\documentclass{aastex62}
\usepackage{mathrsfs}
\usepackage{amsmath}
\usepackage{color}
\usepackage{xcolor}

\begin{document}

\title{Relativistic Tidal Disruption and Nuclear Ignition of White Dwarf Stars by Intermediate Mass Black Holes}

\author{Peter Anninos}
\affil{Lawrence Livermore National Laboratory, Livermore, CA 94550, USA}
\author{P. Chris Fragile}
\affil{Department of Physics and Astronomy, College of Charleston, Charleston, SC 29424, USA }
\affil{Kavli Institute for Theoretical Physics, Santa Barbara, CA, USA}
\author{Samuel S. Olivier}
\affil{Department of Nuclear Engineering, University of California, Berkeley, CA 94720, USA}
\author{Robert Hoffman}
\affil{Lawrence Livermore National Laboratory, Livermore, CA 94550, USA}
\author{Bhupendra Mishra}
\affil{JILA, University of Colorado and National Institute of Standards and Technology, 440 UCB, Boulder, CO 80309-0440, USA}
\affil{Kavli Institute for Theoretical Physics, Santa Barbara, CA, USA}
\author{Karen Camarda}
\affil{Department of Physics and Astronomy, Washburn University, Topeka, KS 66621, USA}

\begin{abstract}
We present results from general relativistic calculations of the tidal disruption of white dwarf
stars from near encounters with intermediate mass black holes. We follow the evolution of
0.2 and $0.6 M_\odot$ stars on parabolic trajectories that approach 
$10^3$ - $10^4 M_\odot$ black holes as close as a few Schwarzschild radii at periapsis, paying particular
attention to the effect tidal disruption has on thermonuclear
reactions and the synthesis of intermediate to heavy ion elements.
These encounters create diverse thermonuclear environments
characteristic of Type I supernovae and capable of
producing both intermediate and heavy mass elements in arbitrary ratios, depending on the strength (or proximity)
of the interaction. Nuclear ignition is triggered in all of our calculations, even at weak tidal strengths
$\beta \sim 2.6$ and large periapsis radius $R_P \sim 28$ Schwarzschild radii. A strong inverse correlation exists between
the mass ratio of calcium to iron group elements and tidal strength, with $\beta \lesssim 5$
producing predominately calcium-rich debris. At these moderate to weak interactions, nucleosynthesis
is not especially efficient, limiting the total mass and outflows of
calcium group elements to $< 15$\% of available
nuclear fuel. Iron group elements however continue to be produced in greater quantity and ratio with increasing tidal
strength, peaking at $\sim 60$\% mass conversion efficiency in our closest encounter cases.
These events generate short bursts of gravitational waves with characteristic frequencies 0.1-0.7 Hz and
strain amplitudes $0.5\times10^{-22}$ - $3.5\times10^{-22}$ at 10 Mpc source distance.
\end{abstract}


\section{Introduction}

Tidal disruptions (TD) of white dwarf (WD) stars by intermediate mass black holes (IMBH) are
complex and violent cosmic events capable of generating significant electromagnetic
and potentially observable gravitational wave energies.  A star passing by a black hole
can be disrupted (elongated and torn apart) while also suffering compression when the black hole tidal
force exceeds the star's self-gravity. The debris from disrupted stars will either dispense unbound materials
into the surrounding medium, or accrete onto the black hole and 
produce flares and jets that can be observed electromagnetically
\citep{Rees88, Haas12, Macleod16}.

Unlike their supermassive counterparts, 
observational evidence for IMBHs in their likely host environments, dwarf galaxies or
globular clusters, is tentative at best \citep{Gerssen02, Gerssen03, Gebhardt02, Gebhardt05, Dong07}, 
fueling efforts to identify
signatures from IMBHs like those that might come from
tidal disruption events (TDE). 
Supermassive black holes (greater than $10^5 ~M_\odot$), such as are observed at the centers of many large galaxies, are not likely to disrupt a WD substantially
before swallowing it entirely. IMBHs, on the other hand, can be effective disruptors, leading to accretion rates up to $\sim 10^4 ~M_\odot$ yr$^{-1}$ \citep{Haas12} (henceforth referred to as HSBL12),
producing strong bursts of radiation useful for detecting these events.

The mass function of IMBHs is uncertain but anticipated to be about
$2\times10^7$ to $4\times10^8$ Gpc$^{-3}$ \citep{Baumgardt04,Macleod16}.
WDs for their part, represent the final stellar
evolution of stars ranging from about 0.07 to 10 solar masses, and are commonly found in spiral galaxies
and globular clusters \citep{Reid05, Gerssen02}.
Less common of course are WD-IMBH interactions, but recent estimates
place expected disruption rates at 500 yr$^{-1}$ Gpc$^{-3}$ \citep{Haas12}.

One especially intriguing aspect of TDEs that concerns this work is the possibility that
tidal compression perpendicular to the orbital plane
could trigger explosive thermonuclear reactions and create heavy-ion nuclei
if the temperature is raised
above $\sim 3\times10^8$ K for He burning or $\sim 3\times10^9$ K for C/O
\citep{Luminet89, Rosswog09, Haas12, Tanikawa17, Kawana17}. 
This is a likely outcome if the WD is massive enough and tidal compression strong
enough. The compression strength in turn is affected by the periapsis radius and black hole mass.

If sufficient amounts of radioactive nuclei, e.g. $^{56}$Ni, are synthesized and dispersed along with the
unbound debris, their decay (through the $^{56}$Ni $\rightarrow$ $^{56}$Co $\rightarrow$ $^{56}$Fe chain, which
releases $\sim 1$ MeV gamma-rays) supplies nuclear energy to the ejecta, much of which is
absorbed and reprocessed as optical/UV photons. The resulting transient emissions
might appear similar to Type Ia supernovae with comparable energy release \citep{Macleod16}.
These optical transients are accompanied by small ejecta mass ($\lesssim 1 M_\odot$) and
fall-back accretion signatures producing flares and jet emissions
that emerge at high X-ray and gamma-ray energies,
similar to gamma-ray bursts, but softer in spectrum and longer in time.
\cite{Shcherbakov13}, for example, suggest that
GRB060218 from the {\it Swift} GRB catalog might be a viable candidate for a TDE involving
a WD and a $10^4 M_\odot$ black hole. 
If confirmed, the appearance of GRB060218 in a dwarf galaxy at a distance of 150 Mpc
would indicate IMBHs are more abundant in the local universe than previously thought,
by about an order of magnitude.

On the other hand, less massive WDs or weaker tidal interactions may lead to low-luminosity
explosions with incomplete radioactive synthesis, or may not explode at all.
Incomplete burning can result in relatively large mass fractions of intermediate mass elements (IMEs)
such as $^{40}$Ca, $^{44}$Ti, and $^{48}$Cr. \cite{Holcomb13} demonstrated how these elements might be a natural
outcome from helium detonations in certain conditions typical of WD-IMBH encounters, with
calcium production generally increasing with decreasing stellar density.
\cite{Sell15} used these results to propose the possibility that the disruption of a light WD
composed mainly of $^{4}$He can be a progenitor of
calcium-rich gap transients. These systems, typically found in the outskirts of known
galaxies, exhibit features similar to Type Ia supernovae, but are faint with
high velocities and large calcium abundances \citep{Kasliwal12}.
Disruption from some subset of TDEs involving white dwarfs might produce large quantities of IMEs,
offering a possible production mechanism for calcium-rich transients.

Observational evidence for the existence of TDE transients
(including heavy iron-group elements, IMEs, fall-back accretion flares) 
remains uncertain, as does the very existence of IMBHs.
Identifying distinguishing features of WD tidal events and providing detailed
estimates of the properties of these transients might help constrain the highly uncertain
nature of the black hole mass function for masses between stellar and supermassive,
and place meaningful constraints on the density of IMBHs in the universe.

In this work we consider the disruption and explosion of
0.2 and 0.6 $M_\odot$ WDs approaching to within a few Schwarzschild radii
of $10^3$ - $10^4 ~M_\odot$ non-rotating IMBHs, performing a series of parameter studies in an effort to
clarify under what conditions thermonuclear reactions are activated in these near
encounter scenarios.
Modeling all of the relevant physics in these systems remains a challenging undertaking,
requiring substantial computational resources and fidelity modeling of not just hydrodynamics, but
also more generally magnetic fields \citep[as recently reported in][]{Guillochon17}, radiation, and nuclear reactions. Near encounters, such as those
considered here, additionally require general relativity. We do not consider magnetic fields or radiation
in this work. We do however include general relativistic hydrodynamics, a full relativistic treatment of the black hole
gravitational field, and a coupled thermonuclear reaction network with self-consistent burn energy feedback.
Nearly all numerical work to date associated with TDEs has relied on
Newtonian hydrodynamics and gravity (in many cases augmented by central potentials approximating
relativistic corrections). Very few incorporate general relativity which is needed
to accurately model ultra-close encounters \citep{Laguna93, Kobayashi04}, but
none of that body of work includes nuclear reactions.
Our goals are to extend previous work \citep[e.g.][]{Rosswog09} (henceforth referred to as RRH09) 
to the general relativistic regime,
scope the strength parameter ranges which result in nuclear burning from close
encounters, calculate the synthesis
of intermediate to heavy mass elements,
and explore disruptions of helium-rich WDs
to address whether such events are viable candidates for calcium-rich gap transients.

We begin in Section \ref{sec:methods} with a brief discussion of our numerical methods, physical models (hydrodynamics, equation
of state, nucleosynthesis), and dynamical mesh strategy needed to approach the high
spatial resolution required for nucleosynthesis, particularly along the vertical scale height.
Initial data for the WD stellar profiles and their locations, trajectories, and tidal
strength parameters for all of the case studies conducted in the course of this work 
are specified in Section \ref{sec:initial}.
Our results follow in Section \ref{sec:results}, and we conclude with a brief summary
in Section \ref{sec:conclusions}.
Unless otherwise noted, standard index notation is used for
labeling spacetime coordinates: repeated indices represent summations,
raising and lowering of indices is done with the 4-metric tensor, and
Latin (Greek) indices run over spatial (4-space) dimensions.

\section{Methods}
\label{sec:methods}

All of the calculations presented in this work use the {\sc Cosmos++} computational astrophysics code
\citep{Anninos05,Fragile12,Fragile14}
to solve the equations of general relativistic hydrodynamics with coupled
thermonuclear reactions and energy generation.  {\sc Cosmos++} is a parallel, multi-dimensional,
multi-physics, object-oriented radiation-magnetohydrodynamic code, written to support adaptive
structured and unstructured meshes, for both Newtonian and general relativistic astrophysical applications.
It was originally written for moving and adaptively refined meshes, and has recently 
been upgraded to accommodate adaptive polynomial order refinement when using 
high-order discontinuous finite elements \citep{Anninos17}. For this work
we utilize the High Resolution Shock Capturing (HRSC) scheme using finite volume representation with
third-order piecewise-parabolic interpolations for the flux reconstructions, and evolved 
using a third-order, low-storage Euler time-stepping scheme. We use a Kerr-Schild
representation for the static (non-rotating) black hole spacetime in Cartesian coordinates.

\subsection{Relativistic Hydrodynamics}
\label{sec:hydro}

Ignoring radiation, magnetic fields, and viscous contributions, none of which are exercised
in this work, the stress energy tensor simplifies to
\begin{equation}
T^{\alpha\beta} =
         \rho h u^\alpha u^\beta 
         + P g^{\alpha\beta} 	~,
\label{eqn:gr_Tmn}
\end{equation}
where
$\rho$ is the fluid mass density,
$h=1+\epsilon/c^2 + P/(\rho c^2)$ is the specific enthalpy,
$c$ is the speed of light,
$u^\alpha = u^0 V^\alpha$ is the contravariant 4-velocity,
$V^\alpha$ is the transport velocity,
$g_{\alpha\beta}$ is the curvature metric, 
and $P$ is the fluid pressure providing closure given an appropriate equation
of state (see Section \ref{sec:eos}).

The four fluid equations
(energy and three components of momentum) are derived
from the conservation of stress energy:
$\nabla_\mu T^\mu_{\ \nu} = \partial_\mu T^\mu_{\ \nu} + 
 \Gamma^\mu_{\alpha\mu} T^\alpha_{\ \nu} - \Gamma^\alpha_{\mu\nu} T^\mu_{\ \alpha} = S_\nu$,
where $\Gamma^\alpha_{\mu\nu}$ are the Christoffel symbols and $S_\nu$
represent arbitrary source terms.
In addition to energy and momentum, we also require an equation
for the conservation of mass
$\nabla_\mu(\rho u^\mu) = \partial_\mathrm{0}(\sqrt{-g} u^0\rho) + \partial_i (\sqrt{-g} u^0 \rho V^i) = 0$.

Expanding out space and time coordinates, the four-divergence
($\nabla_\mu T^\mu_\nu = S_\nu$) of the
mixed index stress tensor is written
\begin{equation}
\partial_\mathrm{0}(\sqrt{-g} T^0_\nu) + \partial_i(\sqrt{-g} T^i_\nu)
    = \sqrt{-g} T^\mu_\sigma \ \Gamma^\sigma_{\mu\nu} + \sqrt{-g} S_\nu ~.
\end{equation}
Defining energy and momentum as
${\cal E} = - \sqrt{-g} {T}^0_0$ and
${\cal S}_j = \sqrt{-g} {T}^0_j$, the equations further take on a traditional transport formulation
\begin{equation}
\partial_\mathrm{0} {\cal E} + \partial_i( - \sqrt{-g} T^i_0)
    = - \sqrt{-g} T^\mu_\sigma \ \Gamma^\sigma_{\mu 0} ~,
\end{equation}
or
\begin{equation}
\partial_\mathrm{0} {\cal E} + \partial_i({\cal E} V^i)
    + \partial_i ( \sqrt{-g} P V^i)
    = - \sqrt{-g} T^\mu_\sigma \ \Gamma^\sigma_{\mu 0} ~,
\end{equation}
for energy, and
\begin{equation}
\partial_\mathrm{0} {\cal S}_j + \partial_i( \sqrt{-g} T^i_j)
    = \sqrt{-g} T^\mu_\sigma \ \Gamma^\sigma_{\mu j} ~,
\end{equation}
or
\begin{equation}
\partial_\mathrm{0} {\cal S}_j + \partial_i({\cal S}_j V^i)
    + \partial_i ( \sqrt{-g} P ~g^0_j~V^i)
    = \sqrt{-g} T^\mu_\sigma \ \Gamma^\sigma_{\mu j} ~.
\end{equation}
for momentum, and
\begin{equation}
\partial_\mathrm{0} {\cal D} + \partial_i ({\cal D} V^i) = 0 ~,
\end{equation}
for mass conservation, where ${\cal D} =\sqrt{-g} u^0 \rho = W\rho$ is the boost density.

Mesh motion is implemented by a straight-forward
replacement of generic advective terms
\begin{equation}
\partial_\mathrm{0} (\sqrt{-g} T^0_\alpha) + \partial_i(\sqrt{-g} T^0_\alpha V^i)
\end{equation}
with
\begin{equation}
\partial_\mathrm{0} (\sqrt{-g} T^0_\alpha) + \partial_i(\sqrt{-g} T^0_\alpha (V^i - V_g^i)) + \sqrt{-g} T^0_\alpha \partial_i V_g^i ~,
\end{equation}
where $V_g^i$ is the grid velocity, and $T^0_\alpha$ is used here to represent any evolved field,
including ${\cal E}$, ${\cal S}_j$, and ${\cal D}$.
The grid velocity can be set arbitrarily by the user, or computed internally
by Lagrangian, softened Lagrangian, or potential relaxing algorithms. At the
end of each cycle, the zone objects (consisting of node positions, oriented face area vectors,
cell volumes, etc.) are physically moved by projecting the zone-centered grid velocity
to the nodes and moving them accordingly to the next time cycle, after which all
local zone object elements (including the spacetime metric) are updated.

At the beginning of each time cycle a series of coupled nonlinear equations
are solved to extract primitive fields (mass density, internal energy, velocity)
from evolved conserved fields (boost density, total energy, momentum), after which
the equation of state is applied to compute the thermodynamic quantities:
pressure, sound speed and temperature.
For Newtonian systems this procedure is straightforward, but relativity
introduces a nonlinear inter-dependency of primitives so
their extraction from conserved fields requires special (Newton-Raphson) iterative treatment.
We have implemented several procedures for doing this, solving
one, two, or five dimensional inversion schemes for magneto-hydrodynamics,
or a nine dimensional fully implicit method (including coupling terms)
when radiation fields are present \citep{Fragile12,Fragile14}.

\subsection{Equation of State}
\label{sec:eos}

We adopt a 2-component equation of state for this work:
\begin{eqnarray}
P &=& P_\mathrm{rad} + P_\mathrm{gas} = a_R T^{4}/3 + (\Gamma-1) Y \rho T ~,  \label{eqn:eos_p}\\
(e - e_\mathrm{cold}) &=& e_\mathrm{rad} + e_\mathrm{gas} = a_R T^{4} +  Y \rho T  \label{eqn:eos_e} ~,
\end{eqnarray}
where thermal radiation and ions both contribute to the energy density ($e \equiv \rho\epsilon$) and pressure ($P$).
The temperature is calculated from the thermal component of the internal energy
density, subtracting off the ``cold'' energy approximated by the initial barotropic equation
of state specifying the initial WD data in the absence of shocks:
$e_\mathrm{cold} = K\rho^\Gamma/(\Gamma-1)$.
Additionally, $Y$ is defined as 
\begin{equation}
Y = \frac{k_b}{m_n}\left(\frac{1}{\mu_W} + N_e\right) \frac{1}{\Gamma-1}	~,
\label{eqn:eosy}
\end{equation}
where $k_b$ is boltzman's constant, $m_n$ is the nucleon mass, 
$\mu_W = 1/\sum_i \rho_i/(\rho~A_i)$ is
the mean molecular weight formed from isotopic densities,
$A_i$ is the atomic weight of isotope $i$,
$N_e$ is the number of electrons per nucleon,
and $\Gamma=5/3$ is the adiabatic index.
The extraction of temperature from equation (\ref{eqn:eos_e}) follows
\cite{Haas12} \citep[see also][]{Lee05}, but we have generalized (\ref{eqn:eos_p}) to account
for radiation pressure. \cite{Haas12} demonstrated (as we have here) this simplified multi-component treatment
capable of reproducing qualitatively similar phase tracks as more complex degenerate
equation of state models (RRH09) for the range and scope of stellar and
interaction parameters considered in this work.
This approximation might adequately capture hydrodynamic behavior, but it is unclear the extent
to which it affects calculations of nuclear burn products, which can be highly
sensitive to temperature. We anticipate in related future work to implement a more accurate
Helmholtz model and assess sensitivities to equation of state choices.

The nonlinear temperature dependence in equation (\ref{eqn:eos_e}) requires Newton
inversion each cycle to solve for the temperature given the fluid density and internal energy.
Partial ionization is approximated by interpolating the number of electrons per nucleon 
$N_e$ in equation (\ref{eqn:eosy}), between zero (no ionization) and 1/2 (full ionization),
over the range of temperatures $T=5\times10^5$ to $T=5\times10^6$ K.
The mean molecular weight $\mu_W$ is updated each cycle from the isotopic fractions
produced after solving the nuclear reaction network discussed in the following section.

\subsection{Nuclear Network}
\label{sec:nucleo}

{\sc Cosmos++} supports a number of nuclear reaction networks, including the 7- and 19-isotope 
$\alpha$-chain and heavy-ion reaction models developed by
\cite{Weaver78}, \cite{Timmes99} and \cite{Timmes00}. There is of course
a trade-off between speed versus accuracy as measured by nuclear energy production when choosing between
these two networks.
\cite{Timmes00} report 30-40\% differences in energy production rates from the two models
across different stellar environments and burn stages.
We find the more accurate 19-isotope model imposes a tolerable increase in computing time, so we
adopt that network for this work. The 19-isotope network solves heavy-ion reactions using a standard
$\alpha$-chain  composed of 13 nuclei ($^{4}$He, $^{12}$C, $^{16}$O, $^{20}$Ne, $^{24}$Mg, $^{28}$Si, $^{32}$S, $^{36}$Ar, $^{40}$Ca, $^{44}$Ti, $^{48}$Cr, $^{52}$Fe,
and $^{56}$Ni), plus additional isotopes ($^{1}$H, $^{3}$He, $^{14}$N, $^{54}$Fe) to 
accommodate some types of hydrogen burning as well as photo-disintegration neutrons and protons.

To account for the advection of reactant species, the hydrodynamic equations
described in Section \ref{sec:hydro} are supplemented with an additional set of mass continuity equations,
one for each isotope $j$ in the network, resulting in the following 
19 partial differential equations that need to be solved each advance cycle:
\begin{equation}
\partial_\mathrm{0}(W\rho_j) + \partial_i(W\rho_j V^i) = \sum_{k,l,m} \rho_m\rho_l R_{lk}(m) - \rho_j\rho_k R_{kl}(j) ~.
\label{eqn:react}
\end{equation}
Equation (\ref{eqn:react}) is written in simplified form explicitly for, but not limited to, binary reactions,
where $R_{lk} (m)$ represents creation rates contributed by species $m$, and
$R_{kl}(j)$ are the corresponding destruction rates for species $j$.
These rate coefficients are highly nonlinear in temperature and span
many orders of magnitude. As a result, the differential equations for the reaction network 
(the ODEs consisting of the right-hand-side of equation (\ref{eqn:react}) after operator
splitting off advection) are stiff and must be solved
with implicit or semi-implicit integration methods. We have implemented a number of solvers and evaluated
their performance in some of our tidal event calculations, including 4th, 5th and 8th order explicit and
semi-implicit adaptive timestep Runge-Kutta schemes, the variable order Bader-Deufelhard method, 
and a 5th order fully implicit method.
The fully implicit solver proved faster and more robust for this class of problems, completing the
3D calculations faster (by a factor of several) and with fewer (by a factor of three)
first-attempt network convergence failures than its closest competitor (Bader-Deufelhard).

Network failures occur when the solver does not converge to a user-specified fractional tolerance
for each species abundance, which we set to $10^{-6}$. When the solver fails to meet this tolerance over
a full hydrodynamic cycle, that particular zone is flagged and switched over
to a sub-cycling method where the time step is systematically reduced until the solver convergences.
The equation of state is re-calculated consistently at each sub-cycle as is the internal energy
production, properly updating the temperature and reaction rates before each new sub-cycled reaction solve.
The local minimum time step needed for proper convergence is recorded and used to set the global timestep
for the next hydrodynamic cycle. The global time step is also constrained to limit relative
energy production to 25\% fractional change. These highly restrictive conditions are necessary
for achieving stable and accurate solutions since nuclear reaction time steps can easily become much smaller
than hydrodynamic time scales.

The specific energy released by nuclear reactions is calculated from
\begin{equation}
\delta \epsilon_\mathrm{nuc} = N_A \sum_j B_j \delta Y_j	~,
\end{equation}
where $N_A$ is Avogadro's number, $B_j$ is the binding energy, and $Y_j$ is the dimensionless
molar abundance of isotope $j$. It is defined in terms of the mass fractions $X_j=\rho_j/\rho$, atomic weights $A_j$
(proton + neutrons), total mass density $\rho$, and number densities $n_j$ as 
$Y_j = X_j/A_j = \rho_j/(\rho A_j) = n_j(\rho N_A)$.
The corresponding internal energy change (accumulated over all subcycled intervals) 
$\delta e = \rho~\delta\epsilon_\mathrm{nuc}$ is then used
to update the relativistic momentum and total energy fields, after also updating
the pressure change $\delta P$ from the equation of state
\begin{eqnarray}
{\cal E}^{n+1}   & = & {\cal E}^{n}   - \left( (\delta e + \delta P) W u_0 + \sqrt{-g} ~\delta P \right) ~, \\
{\cal S}_j^{n+1} & = & {\cal S}_j^{n} +  \left( \delta e + \delta P \right) W u_j	~.
\end{eqnarray}
In this formulation, we have assumed the WD material is sufficiently optically thick (to photons and neutrons),
that released energy can not be radiated away and is instead absorbed instantaneously in cells where it has been emitted.

Performance of the reaction solver and nuclear energy production model have been validated against
the entire suite of problems and code tests proposed by \cite{Timmes00}, in addition to adiabatically
coupled nuclear flows (e.g., standard Big Bang nucleosynthesis). The hydrodynamic coupling and
subcycling procedure are identical to the methods we had previously developed and vetted for chemical reactive networks
with radiative cooling, and have subjected them to numerous tests, including self-similar reactive flows, molecular species
formation in cosmological shocks, and hydrodynamic shock masking.
We have here only substituted nuclear rates for chemical.

\subsection{Mesh Kinetics}
\label{subsec:mesh}

All simulations use a 3D Cartesian grid centered on the WD, covering a spatial domain of $2 R_\mathrm{WD}$ 
along the $z$-axis and $4 R_\mathrm{WD}$ in the $x-y$ orbital plane,
where $R_\mathrm{WD}$ is the radius of the white dwarf. For most of our calculations, the domain is resolved
with three hierarchical grid layers with $96^3$ cells on the base level
and two additional refinement levels (each level doubling the resolution)
for an effective $384^3$ resolution covering the star.  
For a WD with radius $10^9$ cm, this results in an initial uniform cell resolution
($\Delta x, ~\Delta y, ~\Delta z$) of $\sim (2.1, 2.1, 1.1) \times 10^7$ cm
or in terms of scaled dimensions $\sim (R_{S,3}/15, R_{S,3}/15, R_{S,3}/30)$, where
$R_{S,3}$ is the Schwarzschild radius for a $10^3$ $M_\odot$ black hole.
Adaptive mesh refinement (AMR) is used throughout the calculations, refining 
and de-refining every five hydrodynamic cycles. AMR is triggered by the local mass density,
refining where the density exceeds $\rho_\mathrm{max}/20$, de-refining where the density drops below $\rho_\mathrm{max}/200$, where
$\rho_\mathrm{max}$ is the initial maximum cell density across the grid.

Since the TD events considered here have much in common with our
earlier work on the disruption of the G2 cloud orbiting
our galactic center \citep{Anninos12}, we adopt a similar 
approach for implementing a grid velocity to track the trajectory of the WD.
Grid velocity is used to move the base grid at the mean (density weighted) Lagrangian velocity of the star, 
following its orbital motion in the $x-y$ orbital plane while keeping the star centered on the grid.
Adaptive mesh refinement is performed on the moving base mesh. The combination of grid velocity and AMR together
allows us to use fewer refinement levels to achieve high resolution across the stars throughout their evolution.
By comparison, HSBL12 built 8 levels of refinement to achieve comparable resolution.
We note however that we additionally improve on
the vertical scale height resolution as described next.

Because the WD is compressed along the $z$-axis, perpendicular to the orbital plane, 
mesh motion along that direction requires special treatment. 
\cite{Tanikawa17} argued the importance of capturing shocks within the
scale height of the WD to resolve nuclear explosions.
Estimating the scale height $z_\mathrm{min}$ of the WD in the $z$-direction as
$z_\mathrm{min}/R_\mathrm{WD} \sim \beta^{-2/(\Gamma -1)}$ at periapsis,
where $\beta$ is the tidal strength parameter defined in the next section
\citep{Luminet86,Brassart08,Tanikawa17}, we anticipate
a WD with an initial radius of $10^9$ cm
will require a cell resolution of about $10^6$ cm in a tidal field with $\beta=10$. 
We therefore augment the orbital mesh velocity by introducing a converging component
toward the $z=0$ plane with magnitude
\begin{equation}
|V_g^z(t)| = \zeta_2 ~\left| V^z_{L}(t)\right| ~\left|\frac{z(t)}{L_z(t)} \right|^{\zeta_N} ~,
\end{equation}
where $z(t)$ is the time-dependent coordinate,
$L_z(t)$ is the changing vertical length of the grid along the positive $z$-axis
(distance from the orbital plane to the outer boundary),
$\zeta_2 \ge 0$ is a constant multiplier, and
$|V^z_{L}(t)|$ is the average magnitude of the downward-directed $z$-component of the
star's internal Lagrangian velocity.
The factor $|z(t)/L_z(t)|^{\zeta_N}$ smoothly scales back the vertical grid motion
to zero at the midplane, preventing mesh singularities from forming.
Large values of the exponent $\zeta_N>1$ tend to collapse the outer grid edges 
faster than the midplane cells. Fractional values $0 < \zeta_N < 1$ tend to collapse
the midplane zones at a faster rate. We choose $\zeta_N=0.85$ as a reasonable compromise.
The multiplier $\zeta_2$ is initially set to 5 but deactivated (reset to zero) when
the vertical scale height resolution improves ten-fold to $\sim 10^6$ cm
or equivalently $R_{S,3}/300$. 
This prescription allows the innermost and outermost cells to move inward
towards the orbital plane at different velocities, collapsing the
inner zones by a factor of ten, while cutting the distance to the outer $z$-boundary
planes by about two-thirds, depending on specific case parameters.

When the WD passes periapsis, grid motion in the
$x-y$ plane is modified to keep both the star and black hole on the grid
for the remaining evolution to capture flow circulation. This is accomplished by continuing to track
the orbital velocity of the star, while also allowing the grid to stretch according to
\begin{eqnarray}
V_g^x & = & \mathrm{min}(0, v_{L}^{x}(t)) \frac{x_\mathrm{max}(t) - x(t)}{x_\mathrm{max}(t) - x_\mathrm{min}(t)} \\
V_g^y & = & \mathrm{min}(0, v_{L}^{y}(t)) \frac{y_\mathrm{max}(t) - y(t)}{y_\mathrm{max}(t) - y_\mathrm{min}(t)} ~,
\end{eqnarray}
where $x_\mathrm{max/min}(t)$ and $y_\mathrm{max/min}(t)$ are the evolving maximum/minimum $x$ and $y$ grid positions.
This choice of grid velocity holds the upper right edge of the grid fixed to retain
the black hole on the grid, while allowing the lower left edge to stretch with
the tidal debris and approximately track the stellar centroid as it flows 
past and around the black hole.

The boundary conditions on the outer edges of the grid are set to outflow conditions.
All ghost zone quantities are equated to the values of their adjacent internal-zone neighbors, 
except the velocity component normal to the boundary is set to zero if it points onto the grid. Note, this does not necessarily prevent material from coming onto the grid as the parabolic reconstruction of the fluxes may still produce negative or positive values.

\section{Initial Data}
\label{sec:initial}

There are a few important physical quantities that characterize WD-IMBH systems. First the radius of a non-rotating Schwarzschild
black hole, $R_\mathrm{S}= 2GM_\mathrm{BH}/c^2 \approx 3\times10^8 (M_\mathrm{BH}/10^3 M_\odot)$ cm, where $M_\mathrm{BH}$ is the black hole mass.
Second, the tidal radius $R_\mathrm{T}$ at which the tidal force exceeds self-gravity 
\begin{equation}
R_\mathrm{T} \approx 1.2\times10^{10} \left(\frac{R_\mathrm{WD}}{10^9 \mathrm{cm}}\right)
                       \left(\frac{M_\mathrm{BH}}{10^3 M_\odot}\right)^{1/3}
                       \left(\frac{M_\mathrm{WD}}{0.6 M_\odot}\right)^{-1/3}	~\mathrm{cm}	~,
\end{equation}
where $R_\mathrm{WD}$ is the WD radius, and $M_\mathrm{WD}$ is the WD mass.
From these two quantities, together with the periapsis radius $R_\mathrm{P}$, one can define several dimensionless
parameters representing the strength of the tidal encounter $\beta=R_\mathrm{T}/R_\mathrm{P}$, the condition under which the WD
will be swallowed by the black hole without leaving unbound traces $\beta_\mathrm{S}= R_\mathrm{T}/R_\mathrm{S}$, and the condition under which
the BH will swallow a part of the WD in the first passage (the so-called ``WD swallowing the BH limit'')
$\beta_\mathrm{WD}= R_\mathrm{T}/R_\mathrm{WD}$.
Expressions for the latter two quantities take the following aproximate form:
\begin{eqnarray}
\beta_\mathrm{S}    & \approx & 40 \left(\frac{R_\mathrm{WD}}{10^9 \mathrm{cm}}\right)
             \left(\frac{M_\mathrm{BH}}{10^3 M_\odot}\right)^{-2/3}
             \left(\frac{M_\mathrm{WD}}{0.6 M_\odot}\right)^{-1/3}	~, \\
\beta_\mathrm{WD} & \approx & 12 
             \left(\frac{M_\mathrm{BH}}{10^3 M_\odot}\right)^{1/3}
             \left(\frac{M_\mathrm{WD}}{0.6 M_\odot}\right)^{-1/3}	~.
\end{eqnarray}
A WD is tidally disrupted if $\beta \ge 1$, is not swallowed entirely by the BH if $\beta \le \beta_\mathrm{S}$,
and avoids swallowing the BH (or equivalently avoiding the BH swallowing 
part of the WD at first passage) if $\beta \le \beta_\mathrm{WD}$.
Additionally, one can easily derive an expression for the maximum BH mass $M_\mathrm{BH,max}$ above which tidal disruption
is ineffective by setting $\beta_\mathrm{S}=1$ and solving for $M_\mathrm{BH}$
\begin{eqnarray}
M_\mathrm{BH,max} = 2.5\times10^5 \left(\frac{R_\mathrm{WD}}{10^9 \mathrm{cm}}\right)^{3/2}
                           \left(\frac{M_\mathrm{WD}}{0.6 M_\odot}\right)^{-1/2} ~M_\odot	~.
\label{eqn:bhmassmax}
\end{eqnarray}
Equation (\ref{eqn:bhmassmax}) clearly shows
that supermassive BHs (masses greater than $\sim 10^5 M_\odot$) are not good tidal encounter candidates for white dwarfs.

\begin{figure}
\includegraphics[width=0.5\textwidth]{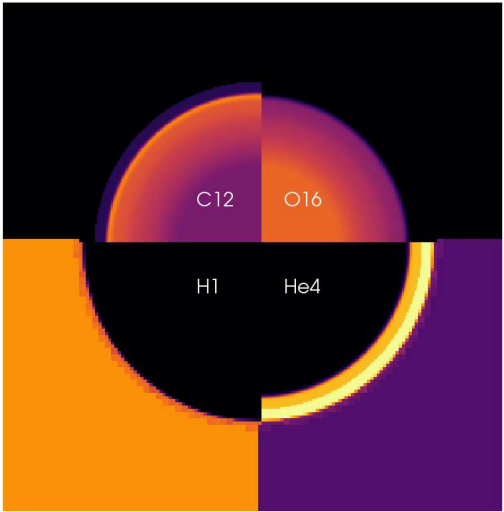}
\caption{
Initial mass fractions of the four dominant isotopes in the interior of the $0.6 M_\odot$ WD.
The four isotopes are layered roughly into four distinct regions: an oxygen-rich center that transitions
to a carbon-rich middle layer, then respectively to thinner helium and hydrogen outer layers.
The background environment is hydrogen/helium gas in primordial abundance.
Color maps are all normalized to the same linear scale from zero (black) to one (yellow).
}
\label{fig:initIsoData}
\end{figure}

Initial data for the stars are derived by solving the relativistic Tolmon-Oppenheimer-Volkof (TOV)
equations with polytropic constant
$K=P\rho^{\Gamma-1} =4.45\times10^{-9}$ (g/cm$^3$)$^{(1-\Gamma)}$ with $\Gamma=5/3$.
These solutions produce $0.6 (0.2) M_\odot$ WDs with radii $1.3 (1.8) \times10^9$ cm, central densities
$7.2\times10^5$ ($1.1\times10^5$) g cm$^{-3}$, central pressures $2.3\times10^{22}$ ($1.0\times10^{21}$) erg cm$^{-3}$,
and total internal energies $3.1\times10^{49}$ ($3.4\times10^{48}$) erg.
We have also computed 0.2 and 0.6 $M_\odot$ WD solutions using the
MESA stellar evolution code \citep{Paxton11} to derive initial isotopics
providing more realistic distributions than the uniform and single (or dual) 
species assumptions often invoked in the literature. For the 0.2 $M_\odot$ case,
a uniform distribution composed mostly of $^{4}$He across the entire radius of
the WD is actually a fair approximation; MESA predicts a 99\% mass fraction of helium,
with trace amounts of hydrogen, carbon, nitrogen, oxygen, neon, and magnesium.
The 0.6 $M_\odot$ WD however exhibits a more complex substructure characterized by
four distinct layers as shown in Figure \ref{fig:initIsoData}. 
The inner half of the WD is roughly a homogeneous
mixture of 1/3 $^{12}$C and 2/3 $^{16}$O (with trace amounts of heavier nuclei), surrounded
by carbon-rich, helium-rich, and hydrogen-rich layers, ordered from the inner core to 
the outer surface.

A few considerations define our choice for the background gas. First, we assume the background
gas is composed of hydrogen and helium in primordial abundance.
Second, we require the mass density be small and not contribute significantly
to the mass accretion rates, but not too low that the
resulting sound speed dominates the compute cycle. We choose $\rho_\mathrm{gas} = 10^{-7} \rho_c$ as a reasonable
compromise, where $\rho_c$ is the initial central density in the WD. 
Third, we require the pressure in the background gas does not produce strong
thermodynamic response from the WD prior to disruption. Large pressure gradients across the stellar
surface can produce bow shocks around the leading surface of the star 
as it moves through the BH environment. Additionally, high background pressures
increase the sound speed and slow the compute cycle.
We set the background pressure to the star surface pressure defined
at $5\times10^{-4}$ of the peak central density.
At these background temperatures and densities, the WD response is dominated by
tidal interactions, not pressure or inertial forces at the surface, even though both the WD and mesh
move supersonically through the background.

All calculations begin with the WD positioned in the $x-y$ orbital plane on parabolic trajectories
defined by periapsis radius $R_\mathrm{P}$ and initial time specified in terms of the
characteristic orbital time $\tau_\mathrm{orb}=2\pi R_\mathrm{P}\sqrt{R_\mathrm{P}/G M_\mathrm{BH}}$ from periapsis at $t=0$. Since all simulations start before periapsis passage, the initial $t$ is negative.
Flexibility in the choice of initial times is used to position stars that are
on different trajectories to comparably scaled distances from the black hole, approximately
one tidal radius so we can reasonably neglect self-gravity of the WD.
Due to the computationally demanding nature of the calculations (particularly when nuclear
reactions are strongly activated, which can easily increase cycle times by factors of a few),
all runs are terminated one to two seconds after the WD passage through periapsis, after
nucleosynthesis completes at energy production levels of
$\delta e_\mathrm{nuc}/(e~\delta t) < 0.01$, but before circularization fully develops.

Although self-gravity is neglected in many of our calculations, we have run several models
with a Cowling approximation to self-gravity, adding the TOV calculated potential as a perturbative
correction to the black hole spacetime metric, centered on the evolving WD centroid. We find our
results are relatively insensitive to this treatment, which is not surprising considering
the contribution to the local gravitational potential is more than an order of magnitude smaller
than the central potential contribution from the black hole at distances within
where the stars are initialized. In addition, the expansion time scale (when self-gravity is neglected)
as estimated by the ratio of the star's diameter to the maximum sound speed is also nearly
an order of magnitude greater than the time interval to travel from unit tidal
radius to where tidal forces are clearly dominant. These conclusions are born out in the
numerical calculations. With the neglect of self-gravity we find stars expand not
more than 10\% from their initial radius. The impact on our final results are equivalently small:
Total nuclear energy production and iron-group abundances differ only by 10\% with and without self-gravity.
Similar behavior is observed when we neglect self-gravity and
instead place the stars at half the tidal radius, in which case there
there is no opportunity for the stars to expand before they are immediately disrupted by the BH.

We have additionally placed stars at twice and three times the temporal tidal distance
from the BH to quantify the impact that early (pre-tidal phase) distortions might have on nucleosynthesis.
Here our findings are less conclusive. The Cowling approximate potential is unable to maintain
precise enough equilibrium through that long a trajectory. The star radius measured along the
vertical direction expands by a considerable amount, more than 20\%, before compression from
the black hole begins to dominate. We thus have two competing forces outside the tidal radius conspiring
to distort the stars: One physical due to the black hole,
one artificial due to the neglect of (or inadequate) self-gravity. Both act to decrease the efficiency
of nuclear activity. We get a sense of the proportion of these two effects by estimating
the time scale for each. Comparing the expansion time $T_{\mathrm{exp}} \sim 2R_{\mathrm{WD}}/c_s$
(where $c_s$ is the peak or central sound speed inside the WD)
against the tidal disruption time $T_{\mathrm{tde}} \sim \sqrt{R_{\mathrm{0}}^3/GM_{BH}}$,
we find they are roughly equal at distances between one to two tidal radii. 
Hence attributing the differences we observe in nuclear activity (both energy and iron-group production)
equally to tidal disruption and thermal expansion, we estimate that the neglect of early-time
distortion affects our results by about 25\%, significantly more than the neglect of
self-gravity inside the tidal radius. We are thus effectively under-predicting
nuclear activity 10\% by neglecting self-gravity inside the tidal radius, but also
over-predicting activity 25\% by neglecting tidal effects outside the tidal radius.
Collectively these two competing effects contribute to an uncertainty of 15\% in our nuclear diagnostics,
not entirely negligible but still nonetheless smaller than the sensitivity we observe with
spatial resolution.

\begin{figure}
\includegraphics[width=0.5\textwidth]{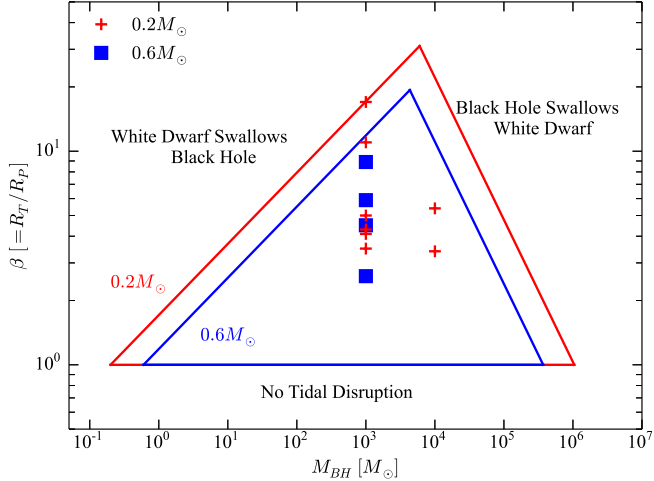}
\caption{Parameter space associated with tidal disruption of $0.2$ (red) and $0.6 M_\odot$ (blue) white dwarfs. Symbols indicate parameter combinations considered in this paper.
}
\label{fig:paramspace}
\end{figure}

Table \ref{tab:runs} and Figure \ref{fig:paramspace} summarize the models we consider in this report, 
exploring three main parameters: black hole mass $M_\mathrm{BH}$, 
white dwarf mass $M_\mathrm{WD}$, and relativistic domain quantified by
the periapsis radius $R_\mathrm{P}$.
Each of these parameters are conveniently encapsulated in the run labels. For example
run B3M2R28 signifies a black hole mass of $10^3 M_\odot$ (B3), a WD mass of 0.2 $M_\odot$ (M2),
and perihelion radius of 28 $R_\mathrm{S}$ (R28).
Also displayed in the table are the
radius of the white dwarf $R_\mathrm{WD}$, initial distance $R_0$ from the
black hole in units of the Schwarzschild and tidal radii, and tidal strength $\beta$.
As mentioned earlier, most of our calculations are performed on a 3-level hierarchical grid with 
a $96^3$ base mesh and two additional
layers of refinement, but we have also run
some calculations at lower resolution (labeled -L2 and -L1 in the table for 2-level and
1-level grids respectively) to assess convergence.
The tidal strength parameter $\beta$ is a derived quantity whose range is restricted by the combination
of black hole mass and $R_\mathrm{P}$. The choice $10^{3-4} M_\odot$ for the black hole mass accommodates
a broad range of $\beta$ as $R_\mathrm{P}$ approaches the Schwarzschild radius.
We do not consider lower mass black holes because we want to 
avoid the $\beta_\mathrm{WD}$ limit where the WD swallows the black hole in near encounters
(essentially keeping $R_\mathrm{WD} < R_\mathrm{P}$ when $R_\mathrm{P} \sim R_\mathrm{S}$).
Higher mass black holes approach the $\beta_\mathrm{S}$ limit, which
results in weaker tidal interactions generally incapable of triggering nucleosynthesis.

\begin{deluxetable}{ccccccc}
\tablecaption{Run Parameters \label{tab:runs}}
\tablewidth{0pt}
\tablehead{
\colhead{Run}         & \colhead{$M_\mathrm{BH}$}          & \colhead{$M_\mathrm{WD}$} 
                      & \colhead{$R_\mathrm{P}$}           & \colhead{$R_\mathrm{WD}$}     
                      & \colhead{$R_0$}                    & \colhead{$\beta$}             \\
                      & ($M_\odot$)                        & ($M_\odot$)
                      & ($R_\mathrm{S}$)                   & ($R_\mathrm{S}$)       
                      & ($R_\mathrm{S}$ [$R_\mathrm{T}$])  & 
}
\startdata
B3M2R28      & $10^3$   & 0.2   & 28 & 6.1   &  98 [0.99]  & 3.5     \\
B3M2R24      & $10^3$   & 0.2   & 24 & 6.1   &  98 [0.99]  & 4.1     \\
B3M2R22      & $10^3$   & 0.2   & 22 & 6.1   &  98 [1.03]  & 4.3     \\
B3M2R20      & $10^3$   & 0.2   & 20 & 6.1   & 104 [1.04]  & 5.0     \\
B3M2R09      & $10^3$   & 0.2   & 9  & 6.1   &  97 [0.97]  & 11      \\
B3M2R06      & $10^3$   & 0.2   & 6  & 6.1   &  93 [0.93]  & 17      \\
B3M6R20      & $10^3$   & 0.6   & 20 & 4.4   &  53 [1.00]  & 2.6     \\
B3M6R12      & $10^3$   & 0.6   & 12 & 4.4   &  49 [1.10]  & 4.5     \\
B3M6R09      & $10^3$   & 0.6   & 8  & 4.4   &  59 [1.10]  & 5.9     \\
B3M6R06      & $10^3$   & 0.6   & 6  & 4.4   &  57 [1.06]  & 8.9     \\
B4M2R06      & $10^4$   & 0.2   & 6  & 0.61  &  21 [1.02]  & 3.4     \\
B4M2R04      & $10^4$   & 0.2   & 4  & 0.61  &  19 [0.90]  & 5.4     \\
\hline
B3M6R20-L2   & $10^3$   & 0.6   & 20 & 4.4   &  53 [1.00]  & 2.6     \\
B3M6R20-L1   & $10^3$   & 0.6   & 20 & 4.4   &  53 [1.00]  & 2.6     \\
B3M6R12-L2   & $10^3$   & 0.6   & 12 & 4.4   &  49 [1.10]  & 4.5     \\
B3M6R12-L1   & $10^3$   & 0.6   & 12 & 4.4   &  49 [1.10]  & 4.5     \\
B3M6R06-L2   & $10^3$   & 0.6   & 6  & 4.4   &  57 [1.06]  & 8.9     \\
B3M6R06-L1   & $10^3$   & 0.6   & 6  & 4.4   &  57 [1.06]  & 8.9     \\
\enddata
\end{deluxetable}

\section{Results}
\label{sec:results}

\subsection{Hydrodynamics}
\label{sec:results_hydro}

\begin{figure}
\includegraphics[width=\textwidth]{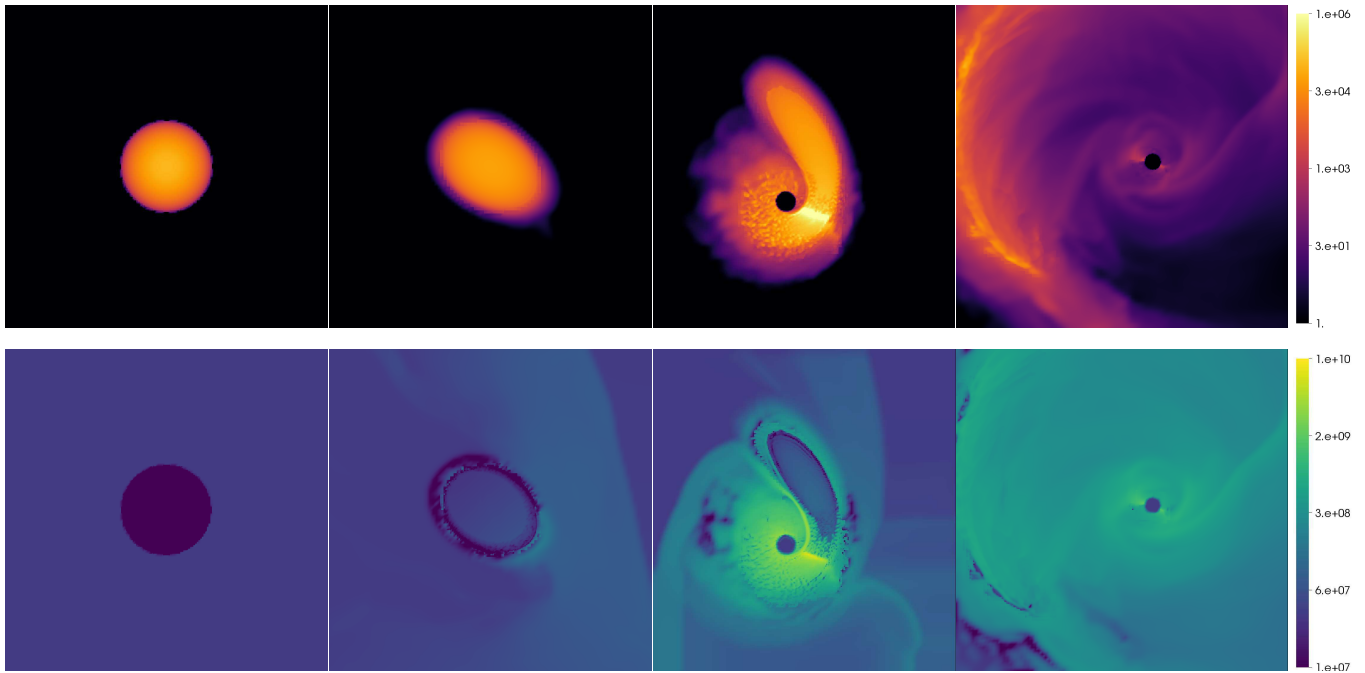}
\caption{
Logarithm of the mass density (in g cm$^{-3}$, top) and temperature (in K, bottom) from the B3M2R06 run plotted in the orbital $x$-$y$ plane 
at $t = -6.4$, -2.0, 0.0, and 2.0 s. The entire cross section of the grid is shown in each frame, but its linear scale changes.
}
\label{fig:tidalxy}
\end{figure}

\begin{figure}
\includegraphics[width=0.5\textwidth]{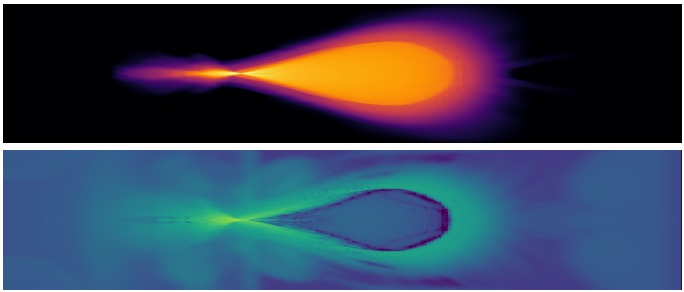}
\caption{
Logarithm of the mass density (in g cm$^{-3}$, top) and temperature (in K, bottom) from the B3M2R06 run plotted in the $y$-$z$ plane near periapsis 
at $t = 0.0$ s. The entire cross section of the grid is shown in each frame, nicely illustrating the star and grid compression in the $z$-direction. The color scale is the same as in Fig. \ref{fig:tidalxy}.
}
\label{fig:tidalyz}
\end{figure}

Several deformation modes are active during typical disruption encounters. In the
$x$-$y$ orbital plane of the binary system, the star will experience tidal stretching along the
direction connecting the black hole to the WD, and tidal compression perpendicular to
this direction. The star will also undergo tidal compression in the direction
perpendicular to the orbital plane. Compression perpendicular to the orbital 
plane is significantly greater than that in the orbital plane, and is the primary
mechanism that could, in principle, ignite nuclear reactions in the WD.
These deformation modes are demonstrated in Figures \ref{fig:tidalxy} and \ref{fig:tidalyz}, which
image the logarithm of mass density and temperature in the $x$-$y$ (orbital) and 
$y$-$z$ planes for run B3M2R06. Stretching and compression in the orbital plane are both evident
in the series of images in Figure \ref{fig:tidalxy}, conspiring to simultaneousy
thin and elongate the WD in orthogonal directions whose orientations change
as the WD approaches the BH. Compression acting
perpendicular to the orbital plane is on display in Figure \ref{fig:tidalyz}, where
the WD collapses to a fraction of its original size. Both figures demonstrate the
effectiveness of the mesh velocity prescription described in Section \ref{subsec:mesh}
to track the orbital trajectory and at the same time collapse the grid in response
to tidal compression along the vertical direction.
Figures \ref{fig:tidalxy} and \ref{fig:tidalyz} are representative of all encounter
scenarios we have studied in that nuclear burning is not triggered and thus has little
effect on stellar disruption until the approach and passage of periapsis.
Ignition is triggered mainly along the dense filamentary-like nozzle structure
that forms where the stellar material is pinched to maximal compression as it passes periapsis, and
is clearly evident in the third plates of Figure \ref{fig:tidalxy} projecting radially outward from the black hole.
Apart from the creation of burn products, the important hydrodynamical effect of nuclear ignition
is the deposition of released energy which contributes to the overall disruption of the WD.

We attribute the small scale structures present in the third panel of 
Figure \ref{fig:tidalxy} to transitory hydrodynamic instabilities.
Material compressing onto the orbital plane before reaching the nozzle is stable, but as Figure \ref{fig:shockyz}
shows, this material bounces after passing through the nozzle and accelerates outward from the
orbital plane, reversing the direction of its vertical velocity and reshocking as it hits infalling
material. In addition, the infalling material is not perfectly aligned with the dense thin shell
of material forming along the nozzle, which creates tangentially aligned streamlines within the dense layer that
can potentially induce perturbations (though we cannot conclusively say this
actually happens due to the lack of sufficient orbital plane resolution).
In all, this creates an interpenetrating, multi-shock morphology inside the disrupted star
that gives rise to interspersed Rayleigh-Taylor unstable regions and
strong shear flows that also trigger Kelvin-Helmholtz instabilities. The result is a complex
flow pattern downstream of the nozzle characterized by debris separation into filamentary structures.
We have verified these features are robust (present) regardless of numerical treatments, including 
activation of AMR, mesh motion, thermonuclear reactions, or equation of state (they are present even with
a simple ideal gas law with no ionization temperature thresholds). 
These features, however, are strongly
sensitive to grid resolution and are less pronounced when the central plane is inadequately resolved.
In fact when we increase resolution, the structures become more coherent, numerous, smaller in scale, and
we expect this trend to continue at resolutions beyond our current limits.
Upon close inspection these features are evident also in the work of other authors (see for example \cite{Kawana17}),
but they are more pronounced in our calculations due to increased spatial resolution and perhaps our use of
less dissipative, high order shock capturing numerical methods.
Although we can say with some degree of confidence these structures are to be expected, clearly
more careful and much higher resolution studies will be needed to better understand their cause and effects.

\begin{figure}
\includegraphics[width=0.5\textwidth]{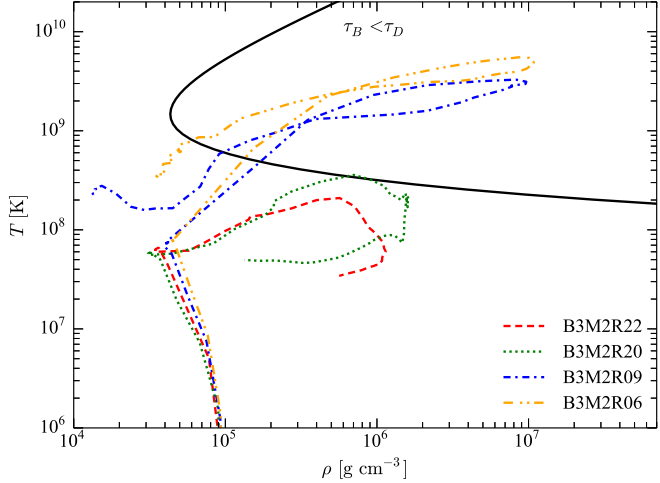}
\caption{
Density-temperature tracks for $0.2 M_\odot$ WD
encounters with a $10^3 M_\odot$ black hole.
The data represent density-weighted averages from the densest 10\% of stellar matter.
The solid black curve is where helium combustion and dynamical reaction time scales
equate. Regions above this curve burn at rates faster than hydrodynamical cooling effects
can suppress nuclear burn. 
}
\label{fig:tracks}
\end{figure}

In general, and as expected, we find more massive WDs compress to greater
densities and temperatures, as do WDs on trajectories approaching closer to
the black hole. The density-temperature phase tracks calculated for the densest 10\% of stellar material and plotted in Figure \ref{fig:tracks}
demonstrate this for a few cases involving encounters with a $10^3 M_\odot$ black hole.
Notice the positive correlation of increasing compression and heating with decreasing perihelion radius.
Also shown in this figure is the curve separating dynamical and helium burn
reaction timescales.
If the dynamical time scale $\tau_D = 1/\sqrt{G \bar{\rho}_\mathrm{WD}}$ is
shorter than the helium combustion timescale
$\tau_B \approx 9\times10^{-4} (T_9)^3 (\rho_6)^{-2} \exp[4.4/(T_9)]$ s
\citep{Khokhlov86}, where $\bar{\rho}_\mathrm{WD}$ is the mean WD density, $T_9 = T/10^9$ K, and $\rho_6 = \rho/10^6$ g cm$^{-3}$,
the star can expand rapidly enough to
quench burning by cooling the WD interior and reducing its density.
Most of the cases we have studied track into the fast combustion timescale regime,
achieving density compression ratios of more than an order of magnitude and temperatures
exceeding those needed to burn carbon-oxygen chains.
The maximum densities and (separately) temperatures observed in our parameter studies are summarized
in Table \ref{tab:results}. The resolution case studies of
B3M6R20, R12, and R06 collectively suggest that the peak densities and temperatures
are converged to about 20\%. This might be adequate if the only
consideration was hydrodynamic behavior or global nuclear energy production (see section \ref{sec:results_nucleo}).
However, nuclear reaction rates are highly sensitive to temperature and we find somewhat greater uncertainty
in the isotopic distributions as discussed in more detail below.

\begin{deluxetable}{cccccc}
\tablecaption{Result Summary \label{tab:results}}
\tablewidth{0pt}
\tablehead{
\colhead{Run}               & \colhead{$\rho_\mathrm{max}$}  & \colhead{$T_\mathrm{max}$}       &
\colhead{$e_\mathrm{nuc}$}  & \colhead{$M_\mathrm{Fe, max}$} & \colhead{$M_\mathrm{Ca, max}$}  \\
                            & (g cm$^{-3}$)                  & (K)                              &
(erg)                       & ($M_\odot$)                    & ($M_\odot$)
}
\startdata
B3M2R28        & $4.5\times10^5$  & $1.4\times10^8$     & $1.3\times10^{47}$  & $<10^{-15}$           & $<10^{-15}$  \\
B3M2R24        & $9.1\times10^5$  & $1.4\times10^8$     & $5.6\times10^{47}$  & $<10^{-15}$           & $1\times10^{-14}$  \\
B3M2R22        & $1.2\times10^6$  & $2.1\times10^8$     & $1.6\times10^{48}$  & $<10^{-15}$           & $9\times10^{-11}$  \\
B3M2R20        & $1.7\times10^6$  & $3.7\times10^8$     & $2.9\times10^{49}$  & 0.0034                & 0.0034  \\
B3M2R09        & $1.1\times10^7$  & $3.4\times10^9$     & $3.1\times10^{50}$  & 0.0891                 & 0.0065  \\
B3M2R06        & $1.2\times10^7$  & $5.8\times10^9$     & $2.8\times10^{50}$  & 0.0798                & 0.0030  \\
B3M6R20        & $2.0\times10^6$  & $5.2\times10^8$     & $1.4\times10^{49}$  & 0.0001              & 0.0002 \\
B3M6R12        & $9.5\times10^6$  & $1.1\times10^9$     & $1.6\times10^{50}$  & 0.0008               & 0.0045  \\
B3M6R09        & $3.0\times10^7$  & $4.1\times10^9$     & $7.6\times10^{50}$  & 0.2646                  & 0.0154   \\
B3M6R06        & $7.2\times10^7$  & $9.0\times10^9$     & $8.7\times10^{50}$  & 0.3690                & 0.0094  \\
B4M2R06        & $3.2\times10^5$  & $2.7\times10^8$     & $7.1\times10^{48}$  & $<10^{-15}$           & $5\times10^{-12}$  \\
B4M2R04        & $6.0\times10^5$  & $1.1\times10^9$     & $1.1\times10^{50}$  & 0.0017                & 0.0306  \\
\hline
B3M6R20-L2     & $1.9\times10^6$  & $4.7\times10^8$     & $1.2\times10^{49}$  & $2\times10^{-15}$     & $8\times10^{-11}$    \\
B3M6R20-L1     & $1.6\times10^6$  & $3.9\times10^8$     & $7.8\times10^{48}$  & $2\times10^{-15}$     & $1\times10^{-12}$  \\
B3M6R12-L2     & $8.2\times10^6$  & $1.7\times10^9$     & $1.7\times10^{50}$  & 0.0016                & 0.0048       \\
B3M6R12-L1     & $6.0\times10^6$  & $1.7\times10^9$     & $1.6\times10^{50}$  & 0.0006                & 0.0018  \\
B3M6R06-L2     & $4.0\times10^7$  & $7.9\times10^9$     & $7.8\times10^{50}$  & 0.3548                & 0.0128  \\
B3M6R06-L1     & $1.8\times10^7$  & $6.5\times10^9$     & $7.1\times10^{50}$  & 0.3171                & 0.0170  \\
\enddata
\end{deluxetable}

\begin{figure}
\includegraphics[width=0.5\textwidth]{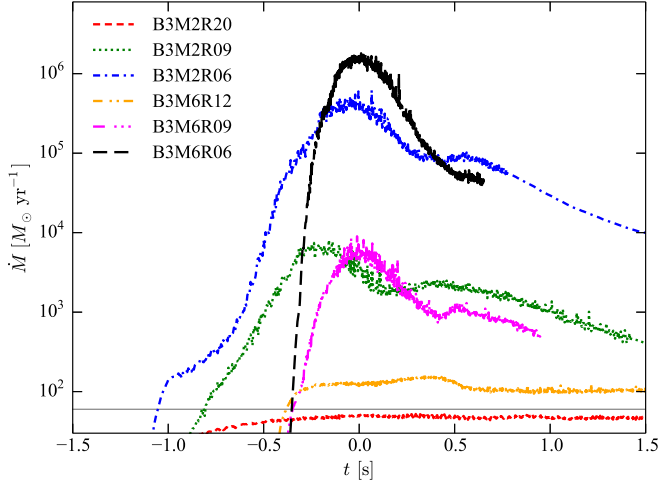}
\caption{
Mass accretion rates for 0.2 and 0.6 $M_\odot$ WD encounters with a $10^3 M_\odot$.
All profiles exhibit similar distinct features: an early large amplitude burst centered at periapsis ($t \sim 0$ s),
followed by a quasi-static phase characterized by a much shallower temporal decay.
The horizontal gray line near the bottom of the plot estimates the background mass accretion rate. 
}
\label{fig:accretionrate}
\end{figure}

Mass accretion rates are plotted as a function of time in Figure \ref{fig:accretionrate} for a sample of cases.
There are two distinct signature regions found in all but the most distant
interaction (largest $R_P$) scenarios: a first burst that peaks for the closest encounters
between $10^6$ - $10^7$ $M_\odot$ yr$^{-1}$ and lasts for 0.5 to 1 second centered at periapsis ($t\sim 0$).
This initial burst is followed by a quasi-static phase with magnitudes in
the range $10^2$ - $10^4$ $M_\odot$ yr$^{-1}$ and a time dependence which, at least for the closest encounter
cases, scales roughly as $\dot{M} \propto 10^{-0.9t}$ as the flow begins to circulate around the black hole
one to two seconds after periapsis.
The range of accretion rates found in the quasi-static phase is considerably
greater than the rate ($\sim 50 M_\odot$ yr$^{-1}$) associated with background gas accretion
represented by the horizontal line in Figure \ref{fig:accretionrate}.
The background accretion rate is highly sensitive to
the background density set in the calculations, but the quasi-static rate is not and appears only to be affected
by grid resolution which dictates coverage of the BH surface boundary and numerical dissipation.
Comparing against purely hydrodynamic calculations, we find nuclear burning enhances mass accretion rates,
total accreted mass, and peak density and temperature tracks by modest amounts, roughly 5-10\%.

These accretion rates are many orders of magnitude greater than the
rates expected from Eddington-limited accretion, which,
assuming a standard 10\% radiative efficiency, implies a rate
\begin{equation}
\dot{M}_\mathrm{Edd} = \frac{L_\mathrm{Edd}}{0.1 c^2} 
               = 4\times10^{-5} \left(\frac{M_\mathrm{BH}}{10^3 M_\odot}\right) M_\odot \mathrm{yr}^{-1}  ~,
\end{equation}
for hydrogen-poor gas with electron scattering opacity
$\kappa_\mathrm{es} = 0.2(1 + X_H) \rightarrow 0.2$ cm$^2$ g$^{-1}$, where $L_\mathrm{Edd} = 4\pi c G M_\mathrm{BH}/\kappa_\mathrm{es}$. 
Assuming a late-time fall-back scaling of $t^{-5/3}$ \citep{Rees88} applied at a return time of
$\approx 10^2$ s, we can reasonably expect these disruption events to emit x-ray and $\gamma$-ray transients and shine at
Eddington luminosity for a year or more.

\subsection{Nucleosynthesis}
\label{sec:results_nucleo}

One of the objectives of this work is to extend the results of RRH09 to the relativistic regime, and provide
independent validation concerning the nucleosynthetics of TDEs using alternative high resolution grid-based numerical techniques
and expanded nuclear reaction networks.
RRH09 used SPH methods coupled with a 7-isotope $\alpha$-network 
to explore a multi-dimensional parameter space of TDEs covering a range of
black hole masses, WD masses, and tidal strengths in search of conditions under which explosive nuclear burning can occur.
They found a generically universal dependence on tidal strength and to a lesser degree on the WD mass.
In particular they found ignition occurs when $\beta \ge 3$ ($\beta \ge 5$) for 1.2 (0.2 and 0.6) solar mass WDs. Their
results also suggest a nominal dependence on the central potential model, generally finding stronger
reactions (greater iron-group production) with a pseudo-potential that accounts for some near-field relativistic effects
compared to purely Newtonian treatments. Hence our interest in exploring fully general relativistic interactions.

Additionally, recent studies by \cite{Tanikawa17} have shed some doubt on the reliability (or limitation)
of multi-dimensional numerical calculations due to the difficulty of allocating sufficient
resources in the central ignition region. They argue coarse resolution does not adequately resolve
shock heating and can produce spurious heating and
artificially trigger ignition. At tidal strengths $\beta \sim 5$,
\cite{Tanikawa17} suggest a spatial resolution of about $10^6$ cm is required to resolve the
scale height sufficiently in order to produce reliable or convergent nuclear reactants.
We have adapted our hybrid AMR/moving mesh technique
to achieve slightly better than the recommended resolution ($\lesssim 10^6$ cm) in the central orbit plane,
improving on previous work by about an order of magnitude.

\begin{figure}
\includegraphics[width=0.5\textwidth]{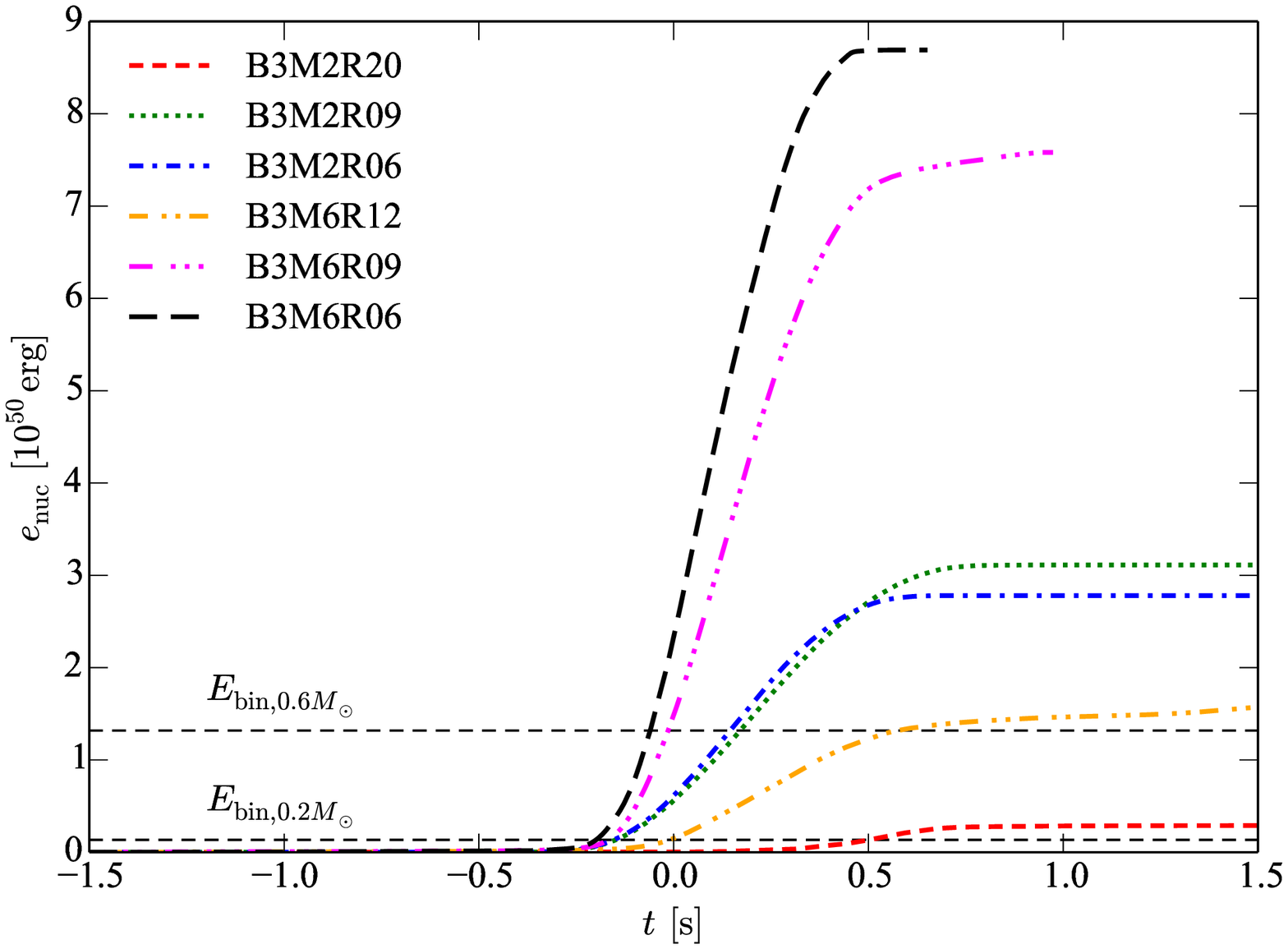}
\includegraphics[width=0.5\textwidth]{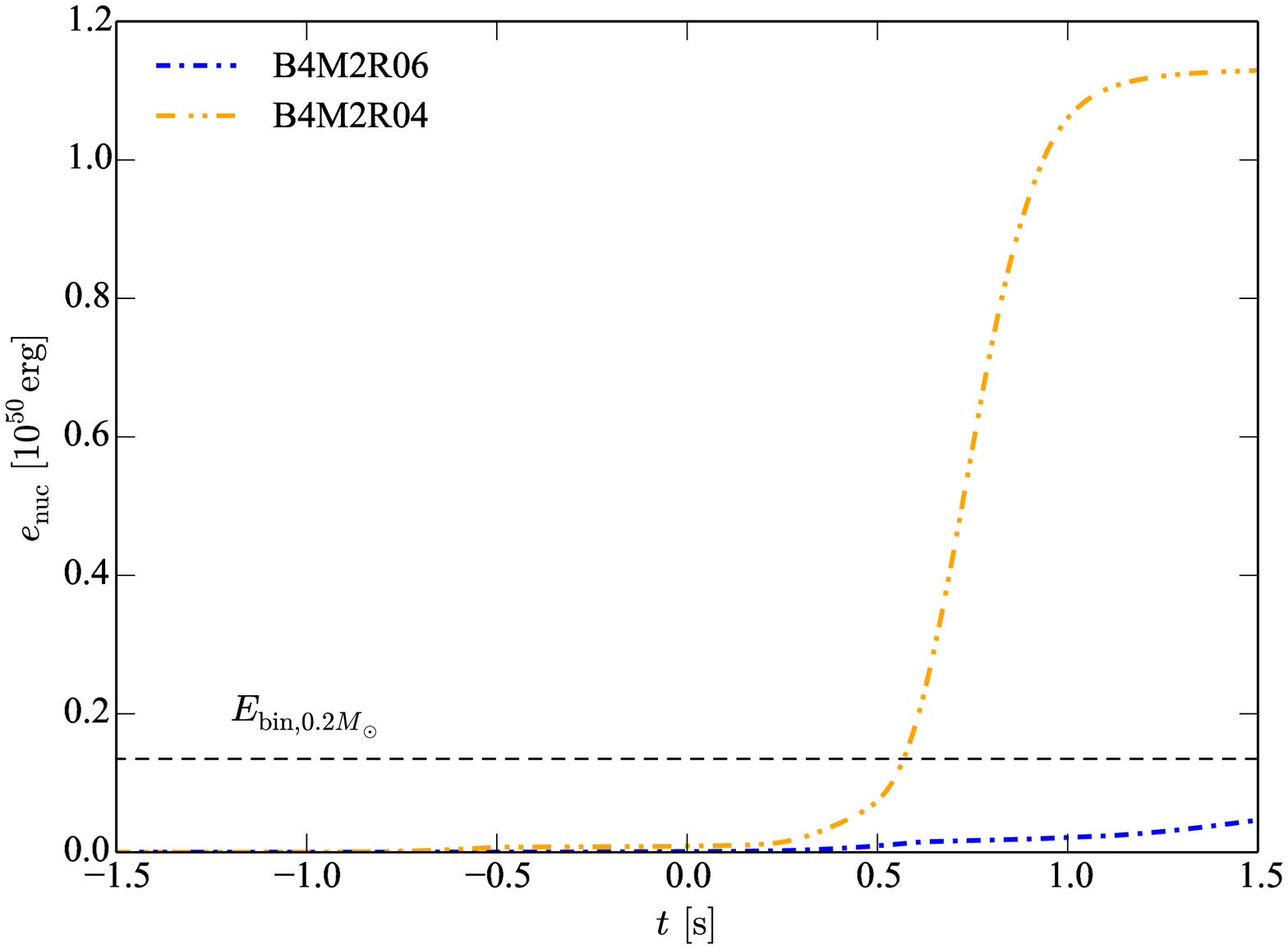}
\caption{Nuclear energy generation for WD encounters with a $10^3 M_\odot$ (left) and $10^4 M_\odot$ (right) black hole.
}
\label{fig:energy}
\end{figure}

A number of diagnostics can be used to quantify
the activation or efficiency of nuclear ignition, including calcium- and iron-group element production and
the total amount of energy released from thermonuclear reactions. We discuss element production in subsequent
paragraphs, but focus first on energy release. The total energy produced by nuclear
reactions is presented in Table \ref{tab:results} for all the cases we considered,
and in Figure \ref{fig:energy} from a select sample. This quantity is not
strongly sensitive to spatial resolution, and is thus a fairly robust measure of thermonuclear ignition.
The trend for energy production is generally monotonically increasing with decreasing $R_P$ (or increasing $\beta$) 
as the WD experiences greater compression, so long as significant chunks are not captured by the BH
to limit the amount of nuclear fuel available for production.
This effect is observed comparing B3M2R09 and B3M2R06. The latter produces slightly less nuclear energy
and iron elements despite achieving higher peak density and temperature.

We find all of our calculations ($\beta \ge 2.6$) ignite, but only those cases
with $\beta \gtrsim 4.5$ release nuclear energies comparable
to the star's binding energies ($1.3\times10^{49}$ and $1.3\times10^{50}$ erg for the 0.2 and $0.6 M_\odot$ WD, respectively).
These results are generally consistent with \cite{Rosswog09}, as are in select (or common) cases the magnitudes of released nuclear energy.
For example, comparing case B3M2R20 to run 6 from RRH09 (with identical tidal strengths $\beta=5$), 
we find $e_\mathrm{nuc}$ agrees to within a factor of two. However, as we point out below, this
level of agreement is not found in all diagnostics.

\begin{figure}
\includegraphics[width=0.5\textwidth]{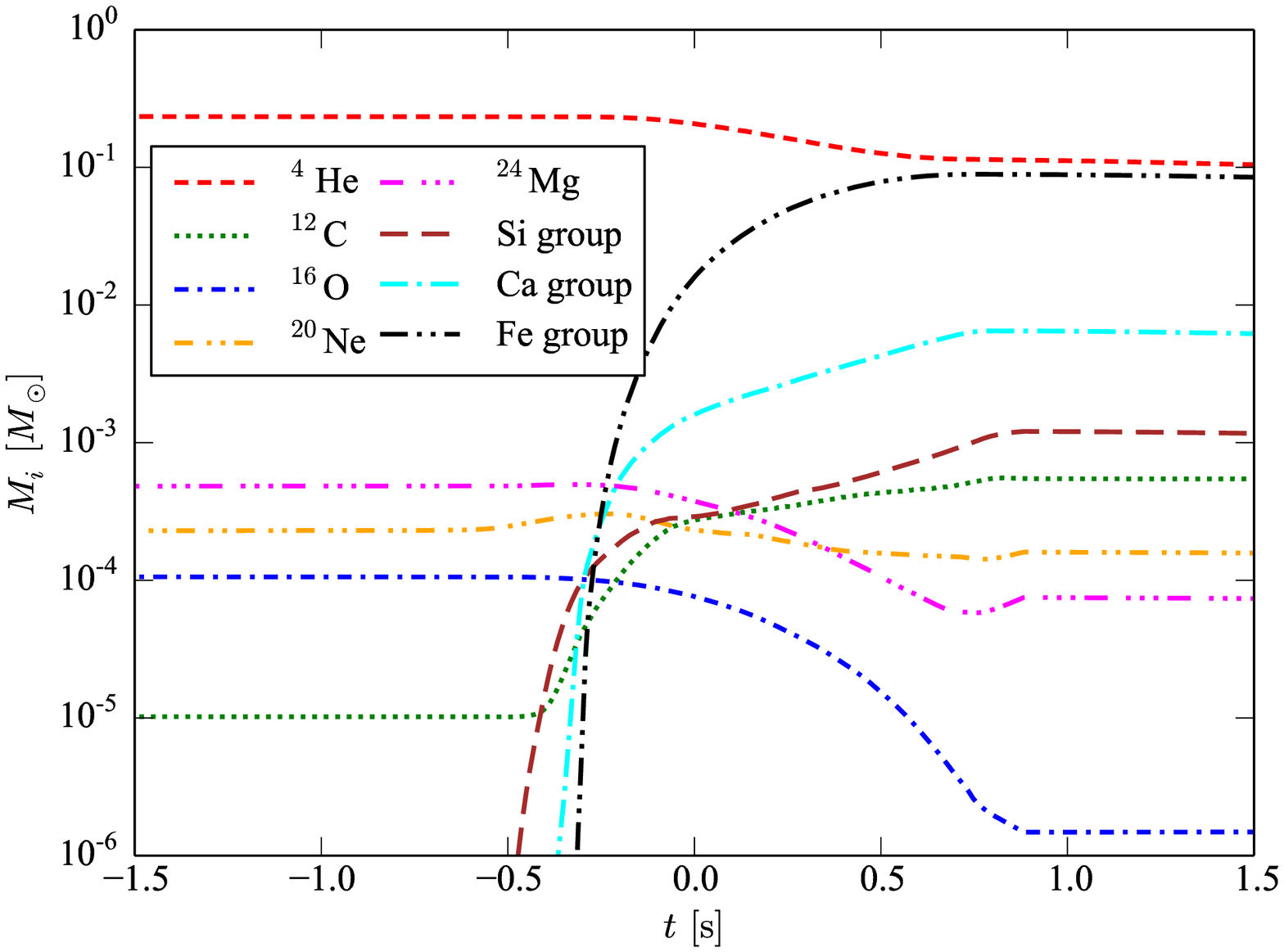}
\includegraphics[width=0.5\textwidth]{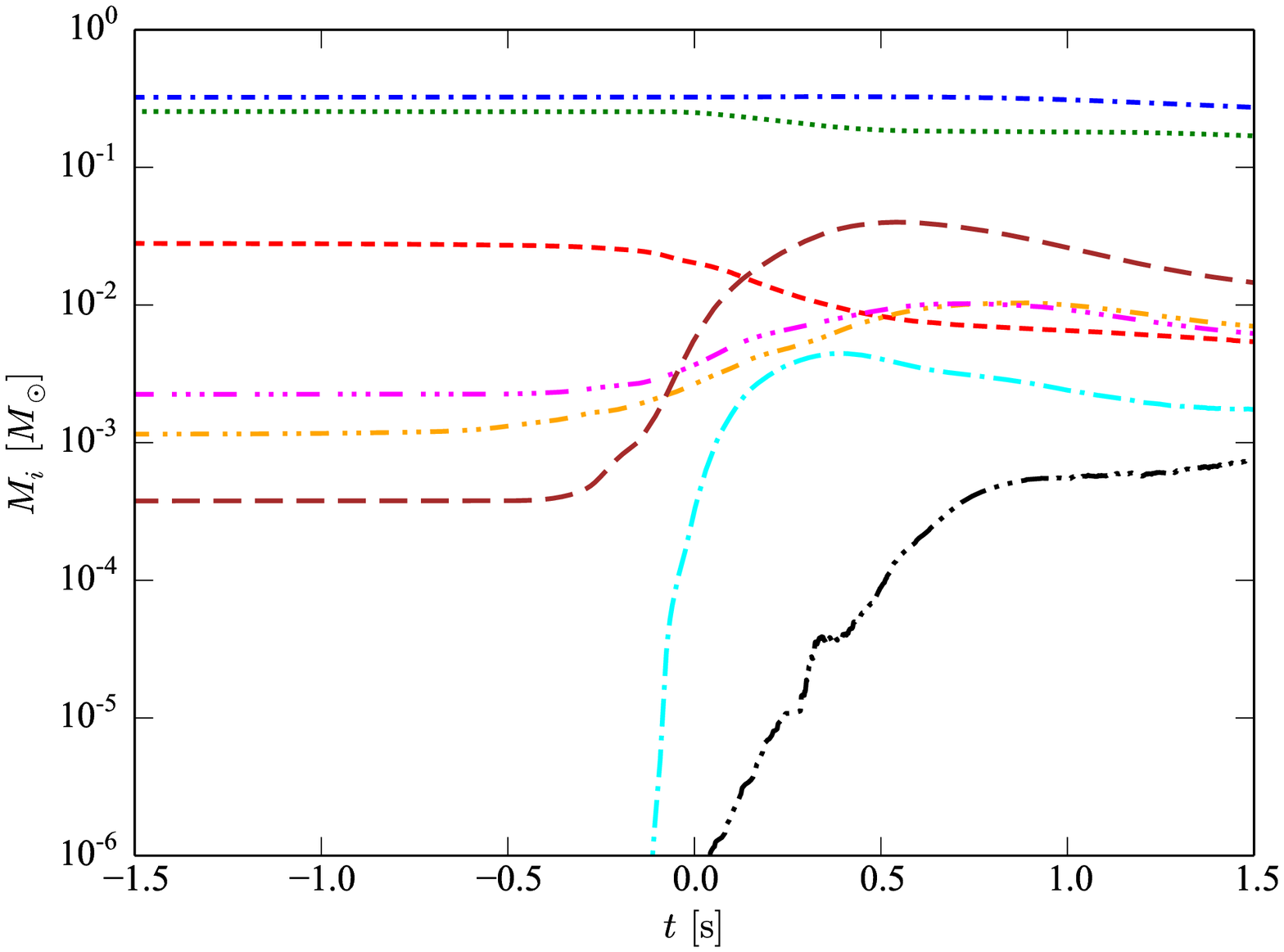}
\caption{Nuclear species mass evolution for simulations B3M2R09 (left) and B3M6R12 (right). The legend is the same for both panels.
}
\label{fig:species}
\end{figure}

An exploding star which releases more energy than its binding energy may not necessarily
produce copious amounts of iron elements. Hence we look separately at the mass of iron-group products
as a complimentary measure of ignition efficiency.
For the 19-isotope network used in our calculations, iron-group is defined
as the sum of iron and nickel isotopes. The peak abundance of iron summed over the entire grid
in each simulation is shown in Table \ref{tab:results}, and plotted as a function
of time in Figure \ref{fig:species} for a couple representative cases along with the other isotopic species.
We additionally introduce silicon- (sum of Si, S and Ar) and calcium- (sum of Ca, Ti, and Cr) group elements to
down-sample and simplify the number of species curves in Figure \ref{fig:species}. A comparison of the 
particular case B3M2R09 (left panel) can be made to Figure 8 in RRH09 by further combining our silicon and calcium
group curves. Although not identical, run 2 in RRH09
and B3M2R09 set the same WD and BH masses and comparable $\beta$ (11 versus 12).
We observe a similar mass deficit for helium elements, an iron abundance that agrees to within
a factor of a few, but very different distributions of all the other minor isotopics.
These differences might be explained by the adoption (or neglect) of trace elements in the initial data
or simplifications necessitated by the different $\alpha$-chain networks (neither the 7 nor 19 isotope networks
can be expected to accurately predict precise distributions, as each makes simplifications to
the reality that would otherwise require modeling hundreds or perhaps thousands of isotopes and reaction paths).
The right panel in Figure \ref{fig:species} plots the species distributions for a moderately disrupted
massive ($0.6 M_\odot$) WD, case B3M6R12. This is an interesting example where the
nuclear flows are choked off before iron production can be completed. In this case, both
calcium and silicon group elements dominate nickel and iron.

\begin{figure}
\includegraphics[width=\textwidth]{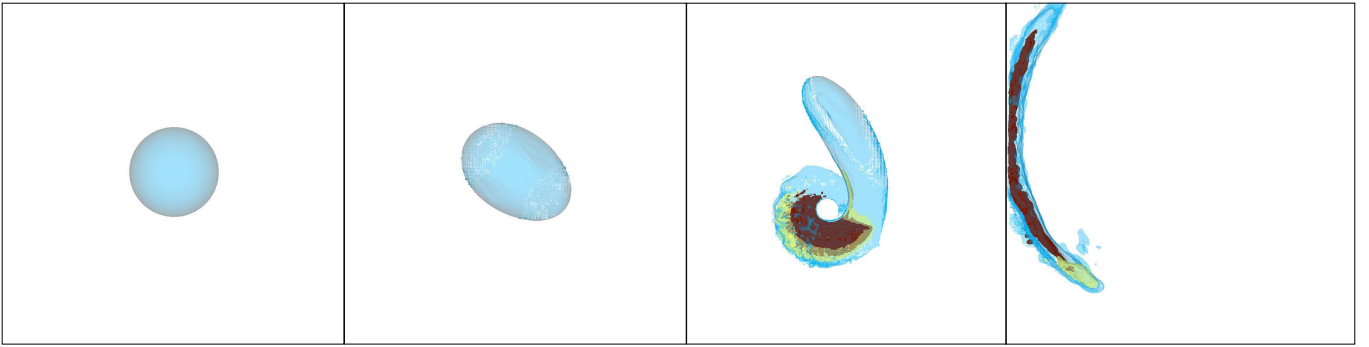}
\caption{
Isosurface plots indicating the distribution of helium (2000 g cm$^{-3}$; light blue), calcium-group (400 g cm$^{-3}$; pale yellow), and iron-group (7000 g cm$^{-3}$; burgundy) from the B3M2R06 run viewed from normal to the orbital $x$-$y$ plane at the same times as in Fig. \ref{fig:tidalxy}.
}
\label{fig:speciesxy}
\end{figure}

Figure \ref{fig:speciesxy} plots a time sequence of three isosurface contours showing the distribution
of helium, Ca-group, and Fe-group elements from the same case (B3M2R06) and times as Figures \ref{fig:tidalxy} and \ref{fig:tidalyz}.
Ca-group elements are first formed when stellar matter is compressed to the proper density and temperature at and around
periapsis, and preferentially along the inner surface closest to the black hole.
Iron group elements are then subsequently formed in their (downstream) wake.
Comparing the third time images in Figures \ref{fig:tidalxy} and \ref{fig:speciesxy} shows nucleosynthesis is initiated 
where the star is maximally compressed along the dense radial filament cut across the star in the
orbital plane as it passes periapsis.
Nuclear products are then subsequently transported in a circular
fashion and mixed alongside unignited stellar and background material around the black hole.

\begin{figure}
\includegraphics[width=1.0\textwidth]{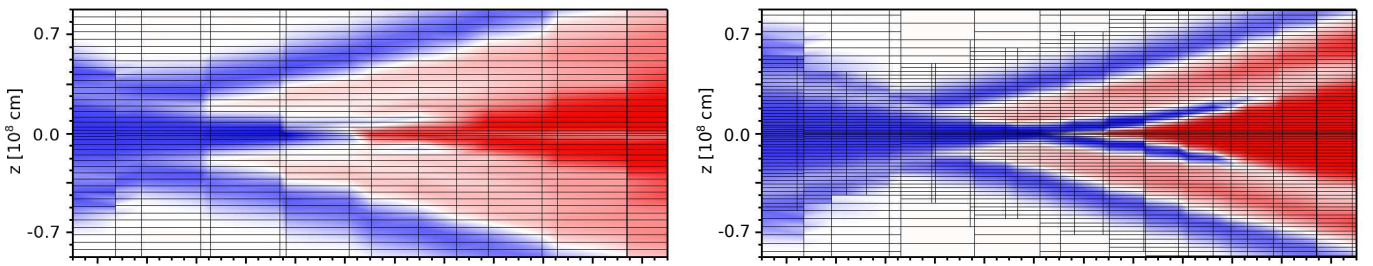}
\caption{Cross section at $t = 0.0$ s of the star centered on (and normal to) the overdense filamentary structure highlighted in the
third plate of Figure \ref{fig:tidalxy}. 
Red colors represent the abundance (mass fraction from 0 to 1) in $^{12}$C+$^{16}$O, while blue colors represent the
abundance of nuclear reaction products (Si+Ca+Fe) from the B3M6R06-L2 (left) and B3M6R06 (right) runs. 
The grid mesh is included to depict how well resolved different features are.
}
\label{fig:burnyz}
\end{figure}

\begin{figure}
\includegraphics[width=0.5\textwidth]{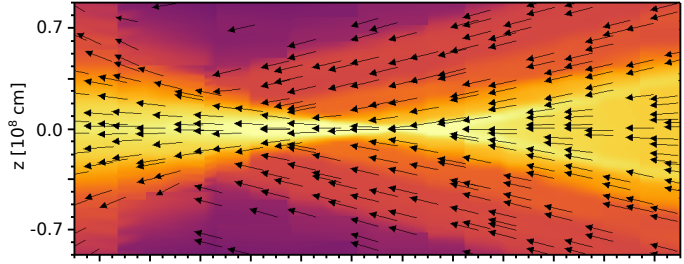}
\caption{Same as right panel of Figure \ref{fig:burnyz}, except the velocity vectors are shown superimposed on the mass density. The density scale is the same as in Figure \ref{fig:tidalxy}.
}
\label{fig:shockyz}
\end{figure}

Figures \ref{fig:burnyz} and \ref{fig:shockyz} present a different orthogonal view of the B3M6R06 event which highlights the complexity
of reactive flows and demonstrates the need for high grid resolution. 
The left and right panels in Figure \ref{fig:burnyz} are taken at the same time ($t=0$)
and represent the same quantities, but are calculated at different grid resolutions: the left panel at L2, the right at L3.
In both images, red represents the mass fraction of combined carbon and oxygen (the initial composition for this event), and blue represents
the combined mass fractions of nuclear products (Si+Ca+Fe groups). Figure \ref{fig:shockyz} shows the mass density (in color) and the
velocity flow (aligned with the arrows) corresponding to the high resolution panel.
In this orientation, the black hole is behind the (zoomed in) stellar matter which is moving
from right to left around the black hole, as it pinches to maximum compression then bounces off the orbital plane. 
The dense radial filament observed in Figure \ref{fig:tidalxy} (though for a different case B3M2R06)
is orthogonal to the image plane and projects towards the black hole at the point of maximum compression located approximately in the middle
of these images (vertically and horizontally). The vertical scale of the image is zoomed in to roughly 6\% of the initial star domain. The
physical length scale between tick marks is the same in both vertical and horizontal axes.
These images complement Figures \ref{fig:speciesxy}, providing a more complete picture of
the disruption process and begin to demonstrate the complex structure of the WD core when it ignites. As noted earlier, nuclear products first form
at and along the pinched feature where density and temperature are highest, responding in time to the
evolving shock structure off the orbital plane producing interpenetrating isotopic
distributions, then transport downstream as they mix with other material.

The total mass of iron-group elements calculated by RRH09 for 0.2 (0.6) $M_\odot$ WDs
ranged anywhere from negligible up to 0.034 (0.0003) $M_\odot$ for the BH masses we have considered, peaking
at tidal strength $\beta=12$ with a 0.2 $M_\odot$ WD, and at $\beta=5$ with a 0.6 $M_\odot$ WD.
Interestingly we find significantly greater iron-group production in our calculations: 0.089 (0.37) $M_\odot$
for 0.2 (0.6) $M_\odot$ WDs at $\beta=$ 11 (8.9). We thus observe three times greater iron production from
0.2 $M_\odot$ WDs, and at least an order of magnitude more from 0.6 $M_\odot$ WDs even after scaling down the tidal
strength to the values quoted in RRH09.

We have additionally considered interactions with large periapsis radii and small to moderate $\beta$ in order to explore
the possibility of producing calcium-rich debris and evaluate whether tidal disruption
events are viable sources of observed calcium-rich gap transients \citep{Sell15}, despite the fact that many of these trajectories are
further from the black hole and do not require full relativistic treatment. 
As Table \ref{tab:results} shows, we have systematically varied parameter sets to approach the low density regime
that is most likely to produce conditions capable of prematurely terminating nuclear reactions
and give rise to dominant calcium group elements (Ca, Ti, Cr) or IMEs rather than iron or nickel. 
\cite{Holcomb13} calculated the smallest critical spatial scale $\xi_\mathrm{crit}$ that is
required for helium detonation at these densities, finding $\xi_\mathrm{crit} \approx 10^9$ ($10^7$) cm
at $\rho \approx 10^5$ ($10^6$) g cm$^{-3}$, well above our grid resolution. They argue incomplete burn occurs if the size of the fuel
reservoir is comparable to $\xi_\mathrm{crit}$, which it is at densities approaching $10^5$ g cm$^{-3}$.
We find, as shown in Table \ref{tab:results}, a direct correlation of systematically
weaker reactions as this density is approached, but a regime clearly exists where Ca group elements dominate over
iron and nickel production before nuclear ignition eventually fails at large $R_\mathrm{P}$.
Table \ref{tab:results} suggests that conditions which give rise to Ca-rich debris are not particularly efficient at nucleosynthesis.
So although calcium-rich outflows can indeed be produced by tidal disruption with IMBHs at low interaction strengths $\beta \lesssim 5$, the
resulting outflows, at least for the encounter parameters we have considered, will consist of
$\lesssim 0.03 M_\odot$ in the form of IMEs, converting roughly 14\% of available fuel mass.

\begin{figure}
\includegraphics[width=0.5\textwidth]{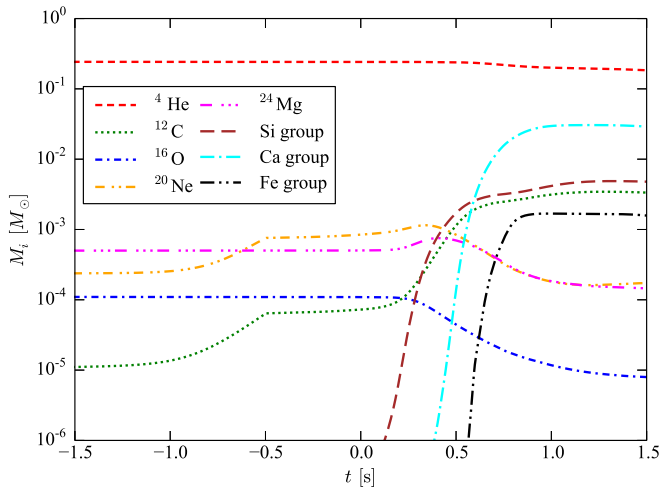}
\caption{Nuclear species mass evolution for simulation B4M2R04.
}
\label{fig:ca_species}
\end{figure}

\begin{figure}
\includegraphics[width=0.5\textwidth]{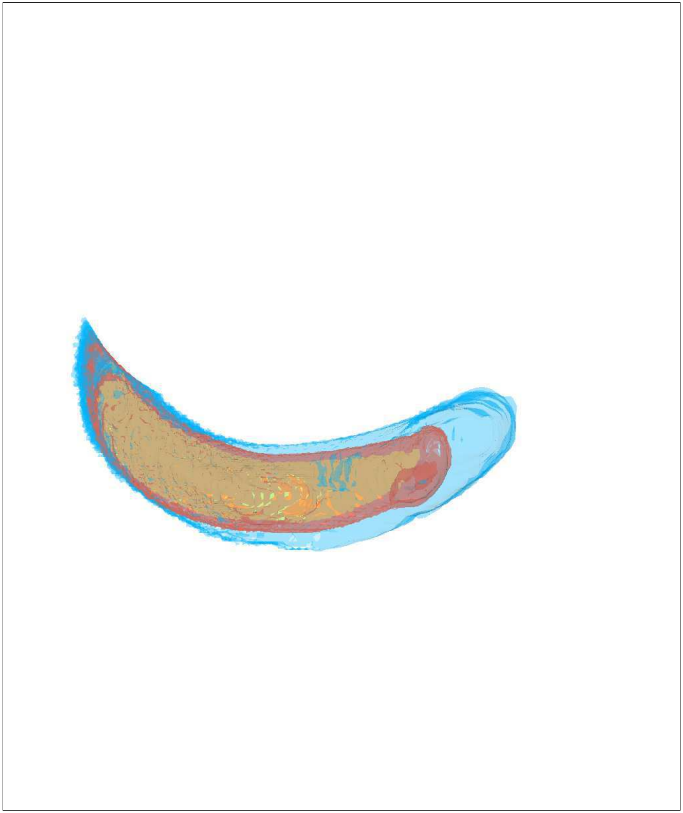}
\caption{
Isosurface plots indicating the distribution of helium (240 g cm$^{-3}$; light blue), silicon-group (1.2 g cm$^{-3}$; orange), and calcium-group (1.2 g cm$^{-3}$; pale yellow) from the B4M2R04 run viewed from normal to the orbital $x$-$y$ plane at time $1.0$ sec.
}
\label{fig:ca_contour}
\end{figure}

In addition to distant encounters with $10^3$ $M_\odot$ BHs, calcium-rich transients may also arise from close encounters
with more massive $10^4$ $M_\odot$ BHs. The range of tidal strengths accommodated by more massive BHs is severely restricted
and consequently less energetic from a nucleosynthesis perspective. Stars cannot be initialized
with periapsis radius closer than about four Schwarzschild radii, or tidal strengths greater than
$\beta \sim 5$, without being promptly captured by the black hole (see Figure \ref{fig:paramspace}).
This is a relatively weak interaction regime where we observe little nuclear activity, so we limit
the number of case studies involving $10^4$ $M_\odot$ BHs to just two calculations.
These two cases, however, prove to be excellent candidates for calcium-rich gap transients precisely because of 
their unique property of sampling both weak tidal strengths and ultra-close encounters. In fact, case B4M2R04
is an example of a moderate strength interaction ($\beta=5.4$) at the closest perihelion radius we have considered, and produces the
greatest mass fraction of calcium-group elements in a calcium-rich environment as demonstrated in Figure \ref{fig:ca_species}.
Figure \ref{fig:ca_contour} shows these transient candidates develop similar layered compositions as their iron-rich counterparts, 
except the dense inner core is composed of calcium-group elements, surrounded by smaller atomic mass elements (e.g., silicon).

\subsection{Post-processed Nucleosynthesis}
\label{sec:results_nucleo_pp}

Table \ref{tab:results} suggests that density-temperature phase tracks are fairly well converged at our highest resolution.
We have therefore extracted time histories of density and temperature for the densest 10\% of matter in the
WD for a few cases, and run a much larger nuclear reaction network
on that point-dimensional data to get a better picture of the full range of element
synthesis in the dense WD core and to compare it against the smaller inline network results. 
The cases we have chosen for post-processing in this way include B3M2R06 and B3M6R06,
the strongest reacting 0.2 and 0.6 $M_\odot$ WDs, respectively.
Because we do not evolve the entire star, but only a sample of its dense core, we simplify the initial isotopic composition
and set it to either pure helium or a 50\% carbon/oxygen mixture for the 0.2
and 0.6 $M_\odot$ WDs, respectively. The data is evolved in time using the Torch code \citep{Timmes99} to solve
stand-alone 19 and 640 isotope reaction networks with time-dependent density and temperature track profiles as inputs, solving the network
system to convergence over each discrete time interval before adjusting density and temperature for the next cycle.
Of course these track histories don't necessarily correspond to contiguous fluid lines, but they are nonetheless useful
for probing reactive transitions in the dense core.

The two cases (B3M6R06 and B3M2R06) are similar in that the 640 network converts more than
90\% of nuclear fuel to iron-group elements, respectively 1.5 and 3 times greater percentage than the global fractions
shown in Table \ref{tab:results}. Global fractions are understandably lower since they sample all stellar material
at very different densities and temperatures beyond the core where iron elements predominately reside.
This is supported by independently post-processing the
dense core data using the same 19 isotope network adopted in the simulations. The 19 and 640 models
predict roughly the same iron-group fraction, differing only by a few percent, giving greater confidence in the
ability of the smaller network to provide reasonable predictions.
Although the end product from the post-processing comes out almost entirely $^{56}$Ni, the 3D simulations predict
84\% (B3M6R06) and 71\% (B3M2R06) of iron-group elements are in the form of $^{56}$Ni, most likely 
attributed to the treatment of photodisintegration.
At the high temperatures achieved in these ultra-close encounters, photodisintegration is very efficient.
Burn products which form while the WD is undergoing initial compression are immediately disassembled
when temperatures approach 6 - 7$\times10^9$ K. Iron-group elements do not
form in substantial quantity until stellar material decompresses and cools to where alpha capture can proceed uninhibited.
Although this effect is observed in some of the species plots from the 3D simulations, it is not as dramatic since that data
represent globally integrated mass fractions and thus sample a range of temperatures at any single time.
It is in these photo-processes where the small and large networks differ the most:
The 19 species model tends to over-estimate photodisintegration and produce more
stable iron isotopes at the expense of nickel. Hence the nickel fractions found in the
3D simulations and quoted above are likely a lower bound to what a larger network might predict.

\begin{figure}
\includegraphics[width=0.5\textwidth]{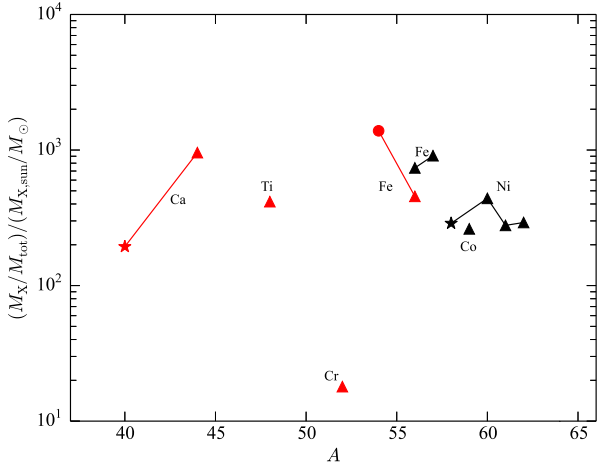}
\caption{Production factor of nuclei produced in the 3D simulation of B3M6R06 (red) and post-processed dense core tracks (black).
Species connected by solid lines are isotopes of the same element (shown), species indicated by a circle are made as
themselves, species indicated by a triangle are made radioactively (and decay to the element shown), species indicated
with a star represent those made as themselves that also have the largest mass fractions in nature.
}
\label{fig:pp_B3M6R06}
\end{figure}

\begin{figure}
\includegraphics[width=0.5\textwidth]{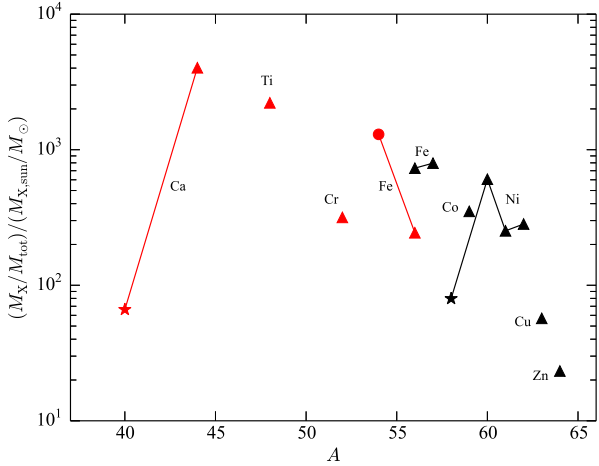}
\caption{As Figure \ref{fig:pp_B3M6R06} but for case B3M2R06.
}
\label{fig:pp_B3M2R06}
\end{figure}

Figures \ref{fig:pp_B3M6R06} and \ref{fig:pp_B3M2R06} compare production factors as a function of atomic mass from the 3D simulations
(red symbols) against the 640-isotope Torch model applied to the dense core tracks (black symbols). All data points represent
the final mass fractions in each run accounting for radioactive decay collapsing all species with a common
mass number to their stable form. These are then normalized to the mass fraction in the sun of the stable species.
Connecting lines indicate isotopes of a given element (listed in the figures), a star indicates the most abundant
isotope of a given element. Circles represent isotopes that were made as themselves, while triangles indicate
that the stable species was made as a radioactive progenitor.
As reactions proceed along the proton-rich side of the alpha chain, they produce stable nuclei up to $^{40}$Ca, after which
unstable nuclei are formed beginning with $^{44}$Ti and produced in our calculations at very high relative (to solar) abundance
as indicated in both figures.
Notice that although calcium represents a small fraction of elements produced in the 3D simulations, it
is comparable to iron and nickel when normalized by solar abundance.
Notice also the extracted phase track results (represented by black symbols)
come out almost entirely as iron-group elements and cannot properly predict
the creation of IMEs forming outside the WD core since they sample only the
densest parts of the WD. This point was made in a previous paragraph, but
emphasized nicely in both Figures \ref{fig:pp_B3M6R06} and \ref{fig:pp_B3M2R06}.
In general the post-processed results suggest conditions similar to Type 1a supernovae and to incomplete silicon burning 
and nuclear statistical equilibrium in Type 2 supernovae: the dominant species produced are $^{56}$Fe and $^{57}$Fe 
made as radioactive nickel, both of which have been observed through $\gamma$-line emission in SN1a and SNII. 
Yet heavier species of cobalt and nickel (made as copper and zinc) are co-produced, but the interesting 
species $^{64}$Zn (made as $^{64}$Ge) is not made in amounts that would suggest tidally disrupted WD's as a possible production site.

Though many details are lost with the 19-isotope model adopted in this work, this exercise demonstrates, among other
things, that it is a fairly good approximation to the overall solution and accurately predicts the dominant end products.
But it is also clear that significant nuclear reactions take place outside the dense
WD core where IMEs are most likely to reside, and that a more accurate post-processing treatment will require
fluid tracers to resolve nucleosynthesis on a global scale across the entire WD.

\subsection{Gravitational Waves}
\label{subsec:gwaves}

Tidal disruption events are potential sources of strong gravitational waves, particularly during the early
phase when the star is sufficiently compact and the interaction highly asymmetric. After passing periapsis,
however, the resulting tidal forces disrupt the star into an extended debris tail and the wave
signal diminishes significantly. Gravitational wave emissions from TDEs will therefore
have burst-like behavior with estimated amplitude (RRH09)
\begin{equation}
h \approx \frac{G M_\mathrm{WD} R_\mathrm{S}}{c^2 R_\mathrm{P} D}
  \approx 1\times10^{-22} \beta \left(\frac{D}{10 ~\mathrm{Mpc}}\right)^{-1}
                                \left(\frac{M_\mathrm{WD}}{0.6 M_\odot}\right)^{4/3}
                                \left(\frac{R_\mathrm{WD}}{10^9 \mathrm{cm}}\right)^{-1}
                                \left(\frac{M_\mathrm{BH}}{10^3 M_\odot}\right)^{2/3} 	~,
\label{eqn:gw_h}
\end{equation}
and frequency
\begin{equation}
f \approx \left(\frac{G M_\mathrm{BH}}{R_\mathrm{P}^3}\right)^{1/2}
  \approx 0.1 ~\beta^{3/2}\left(\frac{M_\mathrm{WD}}{0.6 M_\odot}\right)^{1/2}
                          \left(\frac{R_\mathrm{WD}}{10^9 \mathrm{cm}}\right)^{-3/2} ~\mathrm{Hz}   ~,
\label{eqn:gw_f}
\end{equation}
placing the encounter scenarios we have considered within the anticipated frequency range of LISA \citep{Amaro17}.

\begin{figure}
\includegraphics[width=0.5\textwidth]{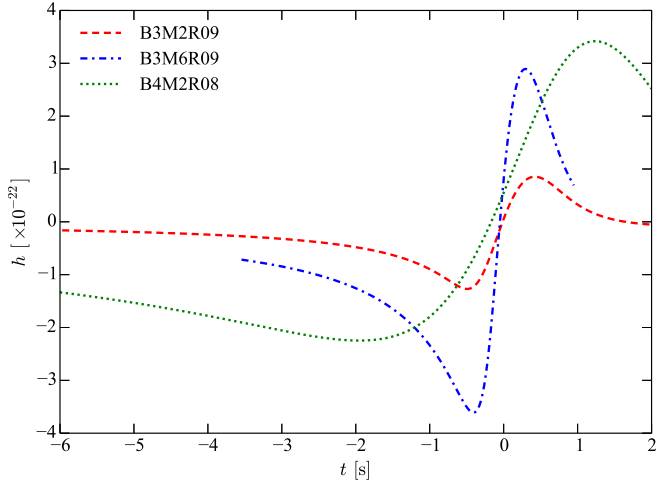}
\caption{Gravitational wave forms ($h_\times$ polarization) for a sampling of simulations with different WD and BH masses, each assuming a source distance of $D=10$ Mpc.
}
\label{fig:gwaves}
\end{figure}

\begin{figure}
\includegraphics[width=0.5\textwidth]{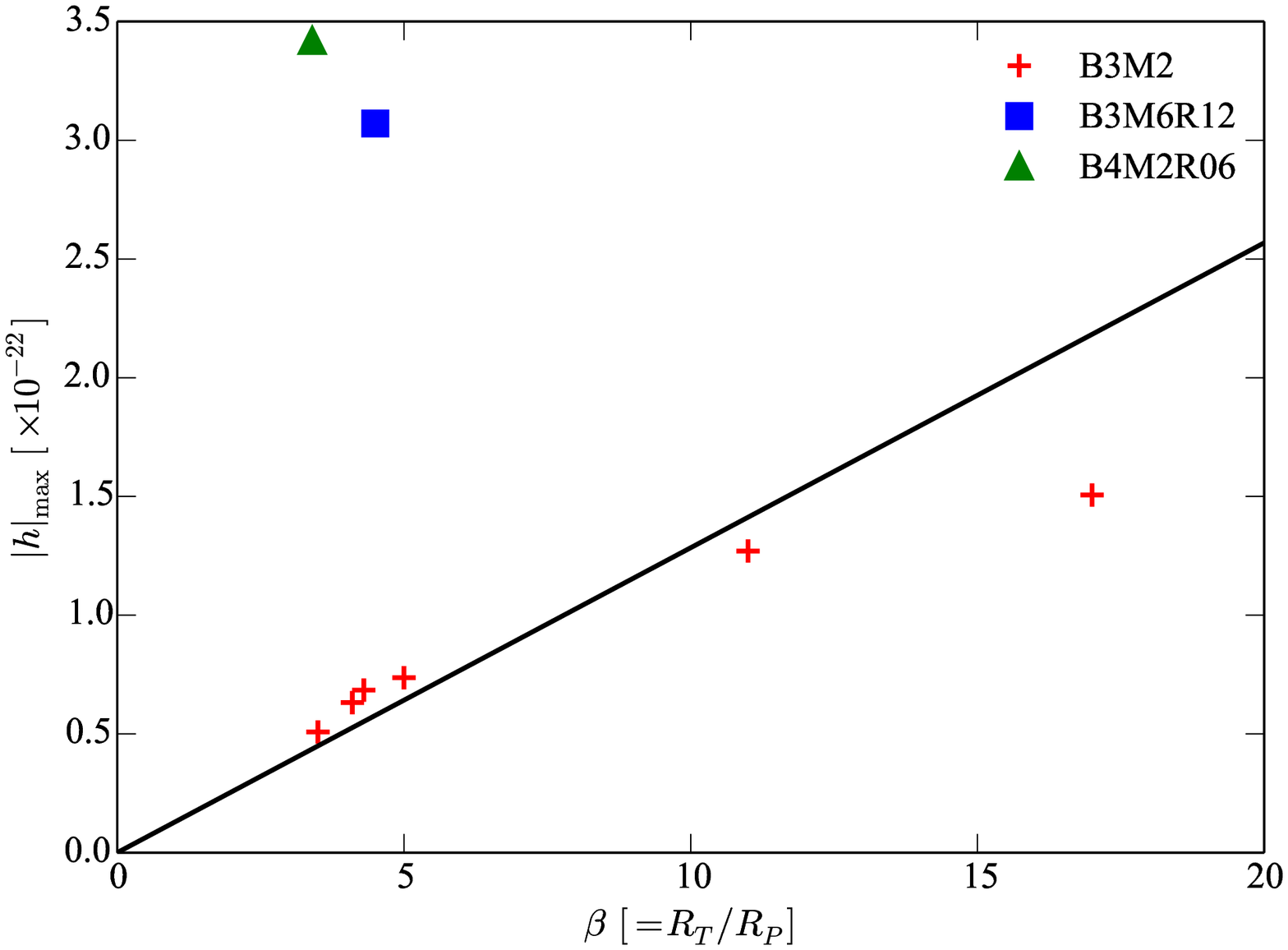}
\includegraphics[width=0.5\textwidth]{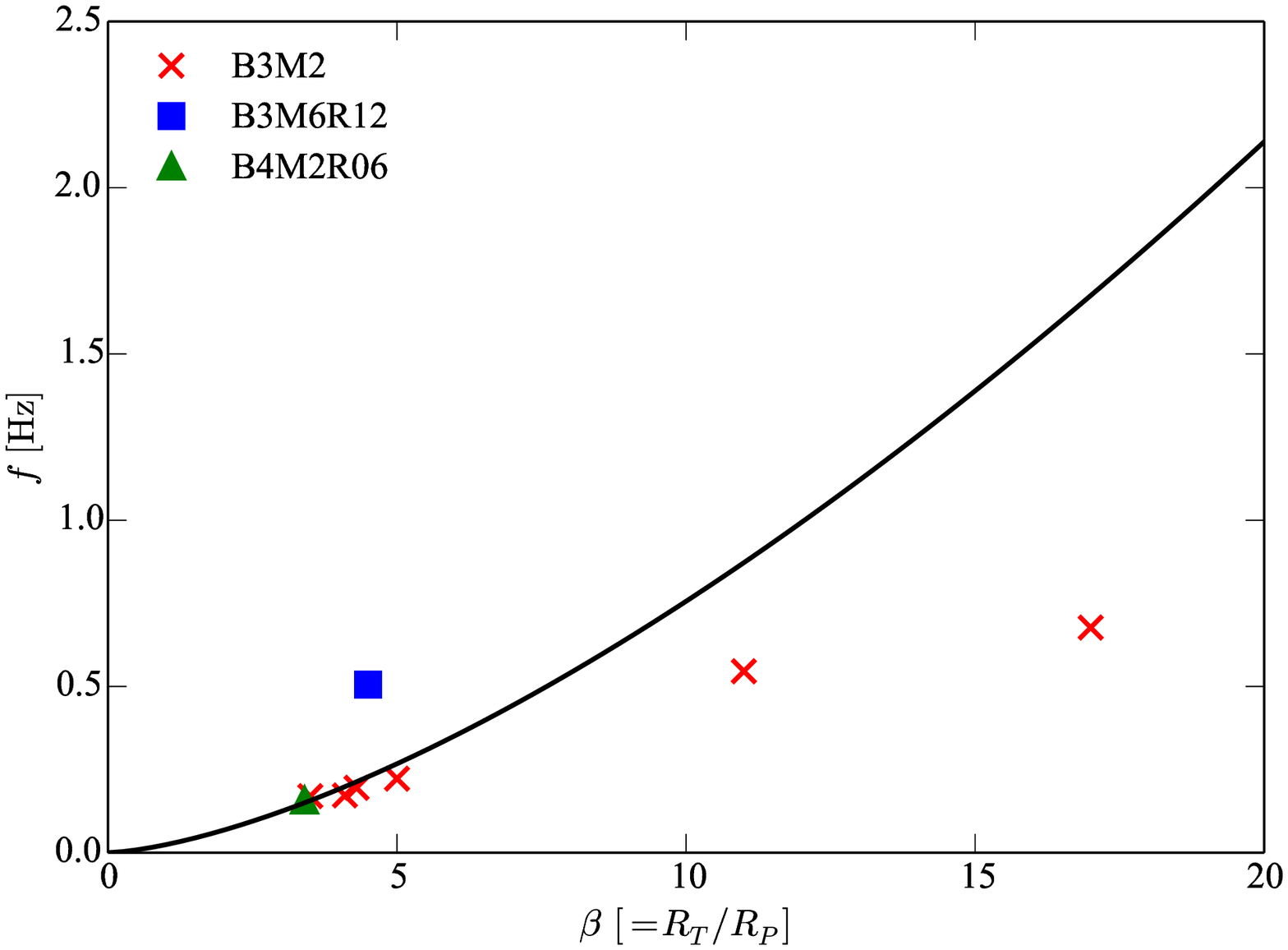}
\caption{Gravitational wave amplitude (at 10 Mpc source distance, left) and frequency (right) as a function of $\beta$. The predicted scalings from 
equations (\ref{eqn:gw_h}) and (\ref{eqn:gw_f}) (solid lines) apply to the B3M2 family of simulations, 
while the other points are included to demonstrate proper scalings with $M_\mathrm{WD}$ and $M_\mathrm{BH}$.
}
\label{fig:gwave_amp}
\end{figure}

More precisely, we calculate the retarded wave amplitudes of both polarizations,
$h_+$ and $h_\times$, for an observer along the $z$-axis at distance $D$
from the source using the quadrupole approximation
\begin{eqnarray}
Dh_+ & = & \frac{G}{c^4} \left(\ddot{I}_{xx} - \ddot{I}_{yy}\right) ~, \\
Dh_\times & = & \frac{2G}{c^4} \ddot{I}_{xy}	~,
\end{eqnarray}
where $\ddot{I}_{ij}$ is the second time derivative of the reduced quadrupole moment of
the mass distribution. In practice, the second time derivatives are eliminated
through substitution of the hydrodynamic evolution equations and integration by parts,
to produce an expression easily calculated from evolved fields \citep{Camarda09}.
We plot in Figure \ref{fig:gwaves} a sample of computed waveforms that showcase their
diversity with different WD and BH mass and interaction combinations.
Figure \ref{fig:gwave_amp} plots the strain amplitude (assuming a 10 Mpc source distance) and frequency for most of our simulations.
The strain amplitude is computed as $\sqrt{h_+^2 + h_\times^2}$, while the frequency is defined simply
as the inverse of twice the time interval between the first minimum and first maximum in the $h_\times$ signal.
The solid lines in Figure \ref{fig:gwave_amp} represent the estimated dependencies predicted by equations (\ref{eqn:gw_h}) and (\ref{eqn:gw_f}).
We find that the $\beta^1$ and $\beta^{3/2}$ analytic scalings for the amplitude and frequency, respectively, 
are, to within small normalization factors, well represented by the data at sufficiently small $\beta$,
but deviate significantly at large tidal strengths ($\beta > 10$).
Although our parameter space does not sample black hole or
white dwarf masses sufficiently to verify their analytic scalings (we considered only two values for each), 
we do note that both strain and frequency scalings
are generally consistent with predictions, increasing with both $M_\mathrm{WD}$ and $M_\mathrm{BH}$.
We also note that the addition of nuclear burning does not significantly affect wave signals. We find,
for instance, wave amplitudes with and without active nucleosynthesis match to within a fraction
of a percent.

Gravitational wave attributes for all events we considered are packed tightly 
between strain amplitudes $0.5\times10^{-22}$ - $3.5\times10^{-22}$ at 10 Mpc source distance, and
frequencies 0.1 - 0.7 Hz, just below and outside the frequency range of LIGO \citep{Abbott16}.
Although these signals fall within LISA's frequency band, the amplitudes are sufficiently small
that they will not likely be observed except at source distances within 10-100 kpc, which,
if current estimates of the WD-IMBH disruption rate (500 yr$^{-1}$ Gpc$^{-3}$) are correct,
would be extremely rare events.

\section{Conclusions}
\label{sec:conclusions}

We have developed a novel numerical methodology based on a combination of moving mesh and adaptive
mesh refinement, and applied it to simulate the tidal disruption of white dwarf stars from near encounters 
with non-rotating, intermediate mass black holes. We solve the general relativistic hydrodynamics equations in a Kerr-Schild
background spacetime, to account for relativistic tidal forces. The AMR/moving mesh hybrid approach
allows us to move the base grid along Lagrangian fluid lines, while at the same time refining on the densest
parts of the white dwarf as it compresses and stretches from tidal forces. This is critical in order to
achieve sufficient zone resources in the central plane regions and provide the necessary scale height coverage to
resolve internal shock and ignition features, which in turn are critical to approach
convergent nuclear reactive solutions. A comprehensive study by \cite{Tanikawa17} suggests a resolution of $\le 10^6$ cm
is needed to properly resolve nuclear flows for moderate tidal strength interactions ($\beta \approx 5$).
Previous three dimensional calculations, (e.g. RRH09, HSBL12) fell short of that standard by about an order
of magnitude. The calculations presented here, to our knowledge, thus represent the highest resolution 3D studies
of these systems to date, achieving vertical resolution of slightly better than $10^6$ cm. However, we caution
that even this resolution is almost certainly inadequate at high interaction strengths.

In addition to improved spatial resolution, this work expands on previous investigations by accounting for
general relativistic hydrodynamics coupled with nuclear reactions composed of a larger isotopic network
and more accurate energy production estimates.
The combination of high resolution and inline nuclear reactions makes these calculations computationally
expensive. Nuclear reactions by themselves increase run times by about a factor of a few 
compared to pure hydrodynamics, depending on local state conditions and reactive response times. As a consequence
our calculations are terminated between one to two seconds after the white dwarfs pass periapsis along parabolic trajectories.
However, this is more than enough time for
nucleosynthesis to complete at 1\% relative energy production levels ($\delta e_\mathrm{nuc}/(e~\delta t) < 0.01$).

Our choice of parameters is motivated by a desire to sample WD mass, BH mass, and
periapsis radius to elucidate the effects of relativity, compression ratios, and
thermodynamic response on nuclear reactions. Options for a 0.2 or 0.6 $M_\odot$ WD brings in
different initial hydrodynamic conditions (density, pressure, temperature), but also
different isotopic compositions (predominately either helium or a carbon-oxygen mixture
initializing isotopic mass fractions with trace amounts of minor isotopics). BH mass of $10^3$ or $10^4 M_\odot$ affects tidal
compression, elongation, and crossing/compression timescales. The periapsis radius modulates relativistic effects
and proximity to the BH. Collectively, these combinations define a tidal
strength parameter $\beta=R_T/R_P$ (ratio of tidal to perihelion radii) which is critically important for predicting
conditions necessary for nuclear ignition. 

We find all of our calculations ignite, even at tidal strengths as low as $\beta =2.6$, creating a wide variety of 
environments capable of converting tidal debris into arbitrary combinations of intermediate and high mass elements 
(or mass ratio of burn products) simply by adjusting the tidal strength parameter.
In a crude sense, considering the debris is far from spherical, stellar matter assembles into
compositions resembling SN Ia models in that the densest parts are composed mostly of iron-group elements 
(or otherwise heaviest possible nuclei depending on the interaction strength), surrounded by IMEs and an outer 
layer of unburned fuel. Stronger interactions (greater $\beta$) of course give rise to systematically
greater energy release and nuclear production of iron group elements.
At tidal strengths $\beta \gtrsim 4.5$ the amount of released nuclear energy
exceeds the binding energy of the stars. This result is consistent with RRH09, as are predictions
of the released energy (to about a factor of two) when we compare runs with comparable interaction strengths.
However, our calculations tend to produce significantly greater amounts of iron group elements,
with differences ranging from factors of three to orders of magnitude.
A peak conversion efficiency, defined by the ratio of iron-group products to the initial stellar mass, 
of about 60\% is observed in our closest encounter scenarios, most of which is in the form of $^{56}$Ni.

These differences are not surprising considering the sensitivity of element production to spatial resolution,
the strong dependence of reaction rates on temperature and therefore the equation of state, and the
limitations of small $\alpha$-chain networks to calculate heavy element production.
It is unclear how much of the observed differences
from previous work is due to improvements made in this paper (general relativity, increased grid resolution, 
and extended network model), to different initial stellar conditions (density, temperature,
isotopic distributions), or to the approximation made for the equation of state and therefore the temperature
that is critical for nucleosynthesis. Any combination of these possibilities could contribute to
the factor of two differences observed in the total energy production diagnostic. However the much greater
differences in iron-group production are not as easily explained and requires further study.

We have additionally investigated whether tidally disrupted WDs might be viable sources of observed
calcium-rich gap transients \citep{Sell15}.
These gap transient systems are characterized by
relatively large calcium abundances compared to iron or nickel, by factors of ten or more, and are thus representative
of incomplete burning environments generally incapable of sustaining chain 
reactions through to iron-group production. Idealized one-dimensional calculations performed by
\cite{Holcomb13} suggest these regimes do exist and might be within the realm of conditions
produced by tidal disruption events, particularly for low mass white dwarfs
at densities between $10^5$ and $10^6$ g cm$^{-3}$ and temperatures $\sim 10^9$ K.
\cite{Kawana17} in their investigations did not observe calcium-rich products, but noted densities found
in their calculations were slightly higher than $10^6$ g cm$^{-3}$.
We point out that conditions appropriate for producing Ca-rich transients are achieved through
moderate to weak tidal interactions, $\beta \lesssim 5$,
where scale height resolution requirements are less demanding and marginally resolved with current methods.
Our investigations show Ca-group elements do dominate over iron and nickel masses at
tidal strengths producing relatively low density cores, with a clear trend for increasing 
calcium to iron abundance ratio with decreasing $\beta$.
However, we also find nuclear reactions at these conditions are inefficient, converting
less than 15\% of available fuel to burn products, the precise composition of which is
a strong function of tidal strength.
So, although we have demonstrated that Ca-rich debris can be produced from the tidal disruption of WDs by IMBHs, the
associated cooler, low density environments
produce corresponding outflows with limited amounts of IMEs ($\lesssim 0.03$ $M_\odot$).

Mass accretion rates are strongly dependent on the interaction strength
and proximity to the BH, but invariably well above Eddington-limited rates. 
We can reasonably expect these interactions to emit x-ray and $\gamma$-ray transients and
shine at Eddington luminosity for more than a year. Rates are generally characterized
by a large prompt burst (for sufficiently close $R_P \lesssim 12$ encounters) lasting between 0.5 to 1 second during periapsis approach, 
followed by a quasi-static phase with a shallower decay profile that scales roughly as $\dot{M} \propto 10^{-0.9t}$
during the onset of the circularization phase. 
At larger perihelion radius, the prompt accretion rate is weak, so the quasi-static phase is
the only distinguishable feature of distant encounters.
The amplitude of the prompt burst is strongly influenced by tidal strength, and peaks between
$10^6$ - $10^7 M_\odot$ yr$^{-1}$ for ultra-close encounters.
Quasi-static accretion rates do not vary as much with tidal strength
and are typically confined to a much smaller and narrower range $10^2$ - $10^4 M_\odot$ yr$^{-1}$ at $R_P \lesssim 12$. 
We find nuclear burning enhances the mass accretion rate, but only by a modest 10\%.

Finally, the encounter scenarios considered here generate burst-like gravitational waves with
amplitudes that are not sensitive to nuclear burning, and
that scale nonlinearly with tidal strength and vary between $0.5\times10^{-22}$ - $3.5\times10^{-22}$
at 10 Mpc source distance. Their characteristic frequencies fall in the range
0.1 - 0.7 Hz, just below LIGO's frequency band but potentially observable with LISA for 
admittedly rare events at source distances within 10-100 kpc.

\begin{acknowledgments}
This work was performed in part under the auspices of the U.S. Department of Energy by Lawrence Livermore National Laboratory under Contract DE-AC52-07NA27344. It used resources from the Extreme Science and Engineering Discovery Environment (XSEDE), which is supported by National Science Foundation grant number ACI-1053575. P.C.F. and B.M. acknowledge support from National Science Foundation grants AST 1616185 and PHY17-48958. B.M. also acknowledges support from NASA Astrophysics Theory Program grants NNX16AI40G and NNX17AK55G and 
NCN grant 2013/08/A/ST9/00795.
\end{acknowledgments}

\software{ Cosmos++ \citep{Anninos05, Anninos12, Anninos17}, MESA \citep{Paxton11}, Torch \citep{Timmes99} }

\end{document}